\documentclass{jpp}
\usepackage{amsmath}
\usepackage{amssymb}
\usepackage{xcolor}
\usepackage{physics}
\usepackage{subcaption}
\usepackage{siunitx}                                
\usepackage{overpic}                                
\usepackage{booktabs}                               
\usepackage{multirow}
\usepackage{hyperref}
\usepackage{upgreek}

\shorttitle{Fluid and kinetic modelling of disruptions}
\shortauthor{I.~Ekmark et al}

\title{Fluid and kinetic studies of tokamak disruptions using Bayesian optimization} 

\author{I.~Ekmark\aff{1}
    \corresp{\email{ida.ekmark@chalmers.se}}, M.~Hoppe\aff{2}, T.~Fülöp\aff{1}, P.~Jansson\aff{3}, L.~Antonsson\aff{1}, O.~Vallhagen\aff{1} \and I.~Pusztai\aff{1}}

\affiliation{\aff{1}Department of Physics, Chalmers University of Technology, G\"{o}teborg, SE-41296, Sweden
    \aff{2} Department of Electrical Engineering, KTH Royal Institute of Technology, Stockholm, SE-11428, Sweden
    \aff{3} Department of Computer Science and Engineering, Chalmers University of Technology and University of Gothenburg, Göteborg, SE-41296, Sweden}

\newcommand{\DREAM}{\textsc{Dream}}
\newcommand{\Lcf}{{\mathcal{L}}}
\newcommand{\Ire}{{I_{\rm re}}}
\newcommand{\Iremax}{{I_{\rm re}^{\rm max}}}
\newcommand{\Irenf}{{I_{\rm re}^{95\,\%}}}
\newcommand{\Irerep}{{I_{\rm re}^{\rm repr}}}
\newcommand{\Iohm}{{I_\Upomega}}
\newcommand{\johm}{{j_\Upomega}}
\newcommand{\jre}{{j_{\rm re}}}
\newcommand{\Iohmfin}{{I_\Upomega^{\rm final}}}
\newcommand{\Ip}{{I_{\rm p}}}
\newcommand{\Ipin}{{I_{\rm p}^{t=0}}}
\newcommand{\taucq}{{\tau_{\rm CQ}}}
\newcommand{\etac}{{\eta_{\rm cond}}}
\newcommand{\nD}{{n_{\rm D}}}
\newcommand{\nNe}{{n_{\rm Ne}}}

\newcommand{\dBB}{{\delta B/B}}

\newcommand{\ncold}{{n_{\rm e}}}

\newcommand{\nre}{{n_{\rm re}}}
\newcommand{\fhot}{{f_{\rm hot}}}
\newcommand{\Tcold}{{T_{\rm e}}}
\newcommand{\me}{{m_{\rm e}}}
\newcommand{\pcI}{p_{\rm c}^{\rm fl}}
\newcommand{\fhotfl}{{f_{\rm hot}^{\rm fl}}}

\begin{document}
    \maketitle

    \begin{abstract}
        When simulating runaway electron dynamics in tokamak disruptions, 
        fluid  models with lower numerical cost are often preferred to 
        more accurate  kinetic models. 
        The aim of this work is to compare fluid and kinetic simulations of 
        a large variety of different disruption scenarios in ITER. 
        We consider both non-activated and activated scenarios; for the 
        latter we derive and implement kinetic sources for the Compton 
        scattering and tritium beta decay runaway electron generation 
        mechanisms in our simulation tool \DREAM~[M. Hoppe et al 2021 Comp. 
        Phys. Commun. 268, 108098]. 
        To achieve a diverse set of disruption scenarios, Bayesian
        optimization is used to explore a range of massive material
        injection densities for deuterium and neon. 
        The cost function is designed to distinguish between successful and 
        unsuccessful disruption mitigation based on the runaway current, 
        current quench time and transported fraction of the heat loss. 
        In the non-activated scenarios, we find that fluid and kinetic 
        disruption simulations can have significantly different runaway 
        electron dynamics, due to an overestimation of the runaway seed 
        by the fluid model. 
        The primary cause of this is that the fluid hot-tail generation 
        model neglects superthermal electron transport losses during the 
        thermal quench. 
        In the activated scenarios, the fluid and kinetic models give 
        similar predictions, which can be explained by the significant influence of the activated
        sources on the RE dynamics and the seed.
    \end{abstract}
    
    \section{Introduction}
    Plasma-terminating disruptions are one of the main challenges facing
    tokamak fusion reactors.
    Disruptions are off-normal events, leading to a sudden cooling of the
    plasma -- a thermal quench (TQ) -- associated with an increase in the
    plasma resistivity, causing the plasma current to decay over the
    longer current quench (CQ).
    The toroidal current cannot change significantly on the short TQ
    time scale, and therefore an inductive electric field is produced,
    which can accelerate electrons to relativistic energies
    \citep{helander2002,Breizman_2019}.
    There are several problems associated with disruptions, including heat
    loads on the plasma facing components and mechanical stresses due to
    electromagnetic forces \citep{hollmann_lims}.
    One of the most significant problems, however, is the generation of
    highly energetic runaway electron (RE) beams.
    
    Runaway generation is expected to be a major problem in the next
    generation of machines with high plasma currents, such as ITER
    \citep{Hender_2007}. 
    The reason is that avalanche multiplication of seed REs is 
    exponentially sensitive to the initial plasma current \citep{avaKin}. 
    In addition, during deuterium-tritium operation, REs can also be
    generated by Compton scattering of $\gamma$ photons and tritium beta 
    decay.
    These {\em activated generation sources} will represent two irreducible 
    sources of energetic electrons and are therefore expected to be the
    dominant RE seed generation mechanisms in future reactor-scale
    tokamaks when hot-tail and Dreicer generation have been successfully
    mitigated.
    An uncontrolled, localized loss of a relativistic runaway beam may result 
    in damage to the first wall.
    One of the most prominent mitigation methods is massive material
    injection (MMI) of a combination of radiating impurities (e.g.~neon)
    and hydrogen isotopes, either in gaseous or solid state.
    Shattered pellet injection is the reference concept for the disruption
    mitigation system in ITER. 
    However, as of yet, the problem of designing a successful disruption
    mitigation strategy is still unsolved.
    
    Several sophisticated simulation codes have been developed in order to
    study disruptions.
    One such code is \DREAM~\citep{dream}, that is primarily developed for
    the study of REs. 
    \DREAM~has several options for simulating electron dynamics -- ranging 
    from fluid to fully kinetic models.
    \DREAM~has already been used to study several aspects of tokamak
    disruption mitigation -- including finding parameter spaces without
    REs for SPARC~\citep{Tinguely_2021,sparcIzzo, sparcTinguely},
    STEP~\citep{esmeeSTEP} and ITER~\citep{bayesJPP}.
    These studies used RE fluid models, i.e.~the momentum-space runaway
    dynamics is replaced by formulae for the steady-state runaway
    generation rates as functions of the background parameters.
    
    Kinetic plasma models are physically more rigorous and can, in certain
    scenarios, yield significantly different results compared with those
    obtained from fluid models.
    In kinetic models, the Dreicer and hot-tail RE generation mechanisms are 
    naturally included because the dynamics in the electron velocity space is 
    modelled by the Fokker-Planck collision operator.
    In certain cases, there may be substantial discrepancies between the 
    kinetic and fluid generation rates.
    For example, in \DREAM, the hot-tail generation mechanism is
    implemented as a generation rate based on a simplified model
    \citep{Smith2008}, which is derived under the assumption that 
    there are no magnetic perturbations. 
    This model can significantly overestimate the RE generation 
    rate when magnetic perturbations are present.
    
    \citet{bayesJPP} used the fluid RE model in \DREAM~to explore the
    injected density space of MMI, with Bayesian optimization, to
    investigate scenarios for disruption mitigation in ITER.
    However, such a comprehensive exploration of the parameter space has 
    not been performed using kinetic modelling.
    Even though isolated studies using kinetic models have been done
    \citep{twoStage, idaPRL}, a thorough exploration of the MMI injected
    density space would not be feasible for a fully kinetic model, due to
    the many simulations needed combined with the high computational cost
    of fully kinetic simulations.
    Using an isotropic kinetic model for such exploration is however 
    feasible, i.e.~the model which evolves a distribution function
    which is analytically averaged over pitch angle.
    Such a procedure is justified for the mildly superthermal region where 
    the complicated hot-tail dynamics takes place, while the relativistic 
    electrons can again be treated in a simplified manner. 
    The isotropic kinetic model in \DREAM~is several orders of magnitude 
    faster than the corresponding fully kinetic model~\citep{dream}.
    
    The aim of this article is to compare fluid and kinetic disruption
    simulations for an ITER-like disruption scenario.
    We explore the injected density space of MMI, using Bayesian
    optimization for sample selection.
    We start by deriving kinetic expressions for the RE generation sources
    from tritium beta decay and Compton scattering (§~\ref{sec:kinSources}).
    Then we proceed by constructing an informative cost function 
    (§~\ref{sec:cost}) for efficient sample selection during the Bayesian
    optimization (§~\ref{sec:simcostbay}).
    
    Our results show that in the non-activated scenarios, the fluid and
    kinetic models yield significantly different optima (§~\ref{sec:res}).
    This can be explained mainly by the generation rate of the runaway seed 
    being overestimated when using the fluid model.
    In the activated scenarios, the fluid and kinetic models yield more 
    similar results. 
    In these cases, the Compton generation mechanism is dominant, which 
    is well represented in the fluid model.
    
    \section{Kinetic runaway electron generation sources} \label{sec:kinSources}
    Runaway electron generation is an inherently kinetic process, as it 
    depends on details in the velocity distribution function that are 
    determined by a balance between collisional friction, electric-field 
    acceleration and radiation processes.
    With the kinetic models in \DREAM, the generation of REs 
    from the Dreicer and hot-tail mechanisms is included in the dynamics
    modelled by the Fokker-Planck collision operator, 
    while REs generated via avalanche multiplication are accounted for using a 
    semi-analytical growth rate by \citet{avalanche} which has been benchmarked 
    against kinetic simulations.

    The activated generation processes, Compton scattering and tritium beta 
    decay, provide energetic electrons that are continually generated over 
    a range of electron velocities. 
    Under the assumption that these generation mechanisms are isotropic in 
    velocity space, kinetic sources can be derived using the same reasoning 
    as for their fluid counterparts \citep{solis}.
    For tritium beta decay, this assumption is justified since the emission 
    angle of electrons during this process is random and uniformly distributed.
    For Compton scattering, gamma photons will be emitted from the walls 
    and enter the plasma from all sides in a nearly isotropic manner, even  
    traversing walls, because the plasma is optically thin to the photons. 
    Therefore the resulting electron distribution can be assumed to be 
    approximately uncorrelated with the pitch angle.

    \subsection{Tritium beta decay} 
    The kinetic source rate due to tritium decay rate $(\dd f/\dd t)_{\rm T}$ can be 
    derived using the expression for the energy spectrum of electrons emitted 
    through beta decay \citep{krane}
    \begin{equation}
        f_\beta(W) \propto F(p,2)p \mathcal{W} (W_{\rm max}-W)^2\quad 
        \text{for }W\le W_{\rm max}\text{, and }f_\beta(W)=0\text{ above} \;W_{\rm max}.
    \end{equation}
    Here $\mathcal{W}=\me c^2\gamma$ is the total energy,  
    $W=\me c^2(\gamma - 1)$ is the kinetic energy,  $\gamma=\sqrt{p^2+1}$ 
    is the Lorentz factor and $p$ is the momentum normalized to $\me c$.
    Since the maximum energy of the emitted electrons is 
    $W_{\rm max}=\SI{18.6}{keV}\ll \me c^2$, we may take the 
    non-relativistic limit of the Fermi function~\citep{fermiBeta},
    \begin{equation}
        F(p, Z_{\rm f})=
        \frac{2\pi\alpha Z_{\rm f}/\beta}
        {1-\exp(-2\pi\alpha Z_{\rm f}/\beta)},
    \end{equation} 
    with the normalized speed $\beta=p/\gamma$, the charge of the final state 
    nucleus $Z_{\rm f}$ (for tritium decay $Z_{\rm f}=2$), and the fine 
    structure constant $\alpha\approx1/137$.
    
    Since the source is isotropic, we may write
    \begin{equation}
        4 \pi p^2 \dd p \left(\dv{f}{t}\right)_{\rm T} = \dd W f_\beta(W).
    \end{equation} 
    Using $\dd W / \dd p=\me c^2\beta$, we find 
    \begin{equation}
        \left(\dv{f}{t}\right)_{\rm T} = 
        \frac{1}{4\pi p^2} f_\beta(W)\me c^2\beta
        \propto \frac{1}{p^2}
        \frac{p \gamma (\gamma_{\rm max}-\gamma)^2}
        {1-\exp[-4\pi\alpha /\beta ]}.
        \label{eq:24}
    \end{equation}
    The requirement that the source integrated over the entire momentum 
    space yields the tritium decay rate, $(\ln 2)\, n_{\rm T}/\tau_{\rm T}$ 
    -- where $\tau_{\rm T}\approx 4500 $ days is the half-life for tritium 
    -- determines the proportionality factor in (\ref{eq:24}) such that 
    \begin{equation}
        \left(\dv{f}{t}\right)_{\rm T}\approx 
        C\frac{\ln2}{4\pi} \frac{n_{\rm T}}{\tau_{\rm T}}\frac{1}{p^2}
        \frac{p \gamma \left(\gamma_{\rm max}-\gamma\right)^2}
        {1-\exp(-4\pi\alpha/\beta)},
        \label{eq:dfTdt}
    \end{equation}
    with the constant
    \begin{equation}
        C = 1\left/
        \int_0^{p_{\rm max}}\frac{1}{p^2}
        \frac{p \gamma \left(\gamma_{\rm max}-\gamma\right)^2}
        {1-\exp(-4\pi\alpha/\beta)} p^2\dd p\right.\approx 31800,
    \end{equation}
    where $p_{\rm max}$ is the momentum corresponding to $W_{\rm max}$.

    \subsection{Compton scattering rate}    
    The kinetic Compton scattering rate can be calculated using the 
    Compton cross-section and the gamma energy flux spectrum emitted by 
    a given wall material.
    When using the kinematic relation between the photon energy 
    $W_\gamma$, and the energy $W$ and deflection angle $\theta$ of the 
    scattered electron, we can write
    \begin{equation}
        \cos\theta=1-\frac{\me c^2}{W_\gamma}\frac{W}{W_\gamma'},
        \label{eq:costheta}
    \end{equation}
    where $W_\gamma' = W_\gamma - W$ is the scattered photon energy.
    The Compton scattering process is described by the Klein-Nishina
    differential cross-section
    \begin{equation}
        \dv{\sigma}{\Omega}=
        \frac{r_{\rm e}^2}{2} \frac{W_\gamma'^2}{W_\gamma^2} 
        \left[\frac{W_\gamma}{W_\gamma'}+\frac{W_\gamma'}{W_\gamma}
        -\sin^2(\theta)\right],
    \end{equation}
    with $r_{\rm e}=e^2/4\pi \varepsilon_0m_{\rm e}c^2$ being the classical 
    electron radius. 

    Thus, in a plasma with the total electron density $n_{\rm e,tot}$, i.e.~the 
    electron population of both free and bound electrons, and isotropic photon 
    distribution, we can write
    \begin{equation}
        4 \pi p^2 \dd p \left(\dv{f}{t}\right)_{\rm C} 
        = n_{\rm e,tot} \int_{W_{\gamma0}}^\infty \dd W_\gamma \left[\Gamma_\gamma(W_\gamma)
        \left(2\pi\sin\theta \dv{\sigma}{\Omega} \right) \dd\theta(W,W_\gamma)\right],
    \end{equation} 
    where $\Gamma_\gamma(W_\gamma)$ is the gamma energy flux spectrum.
    The lower integration limit $W_{\gamma0}$ can be derived from 
    equation \eqref{eq:costheta} as $W_{\gamma0}=(p+\gamma-1)/2$.
    
    \citet{solis} estimated the photon flux energy spectrum in ITER by
    $\Gamma_\gamma(W_\gamma)=\Gamma_0\exp[-\exp(-z)-z+1]$, 
    where ${z = [\ln(W_\gamma/10^6)+1.2]/0.8}$ and
    \begin{equation}
        \Gamma_0=\Gamma_{\rm flux}\displaystyle \left/
        \int_0^\infty \exp[-\exp(-z)-z+1]\, \dd W_\gamma\right. 
        =\Gamma_{\rm flux}/5.8844,
    \end{equation}
    with the total flux ${\Gamma_{\rm flux}=10^{18}}/(\rm m^2 s)$. 
    This applies for an H-mode discharge at 15 MA and 500 MW fusion power. 
    For an L-mode case, when the plasma energy is lower, the total flux 
    would be approximately \SI{25}{\percent} of that of the H-mode \citep{solis}. 
    Part of the flux is caused by the time-integrated activation of the walls, 
    and the typical ratio between the prompt gamma flux from fusion neutrons and 
    the gamma flux after the fusion reactions cease is $1000$. 
    These calculations were performed for a beryllium first wall. 
    ITER currently plans to have a first wall made of tungsten, which might 
    change the photon flux and spectrum.
    
    The runaway production rate from Compton scattering can be obtained as 
    \begin{subequations}
        \begin{align}
            \left( \dv{f}{t} \right)_{\rm C} 
            &=\frac{n_{\rm e,tot}}{2}\frac{1}{p^2}\int_{W_{\gamma0}}^\infty\dd W_\gamma 
            \Gamma_\gamma(W_\gamma) 
            \sin\theta \dv{\sigma}{\Omega}\dv{\theta}{p}
            \label{eq:ComptDiff}\\
            &=\frac{n_{\rm e,tot}}{2}\frac{1}{p^2}\int_{W_{\gamma0}}^\infty\dd W_\gamma
            \Gamma_\gamma(W_\gamma)\dv{\sigma}{\Omega}
            \frac{\beta}{\left(\frac{W_\gamma}{\me c^2}+1-\gamma\right)^2},
            \label{eq:dfCdt}
        \end{align}
    \end{subequations}
    where we have used 
    \begin{align}
        \dv{\theta}{p}=
        \dv{}{p} \arccos\left(1 -
        \frac{\me c^2/W_\gamma}{\frac{W_\gamma}{\me c^2(\gamma-1)}-1} \right) 
        =\frac{p / \gamma}
        {\sqrt{1-\cos^2\theta}\left(\gamma-1\right)^2
            \left(\frac{W_\gamma/\me c^2}{\gamma-1}-1\right)^2}.
    \end{align}

    Both the kinetic tritium and Compton sources are consistent with the 
    fluid runaway generation rate \citep{solis}
    \begin{equation}
        \left(\dv{n_{\rm re}}{t}\right)_{\rm T/C}=
        \int_{p>p_{\rm c}}\dd^3\boldsymbol{p}\left(\dv{f}{t}\right)_{\rm T/C},
        \label{eq:solisTritium}
    \end{equation}
    when the runaway region is set to be above the critical momentum for RE 
    generation $p_{\rm c}$.

    \section{Bayesian optimization of simulated disruptions}
    \label{sec:simcostbay} 
    In this paper, we will compare disruption simulations using an isotropic 
    reduced kinetic model, described in Appendix~B.2 of \citep{dream}, and a 
    computationally less expensive fluid model.
    The comparison of fluid and kinetic simulations for a 
    wide variety of disruption scenarios was facilitated using Bayesian 
    optimization.
    The disruption model and simulation details used are described in 
    §~\ref{sec:simset}, while the Bayesian optimization is detailed in 
    §~\ref{sec:bayesopt}.
    In §~\ref{sec:cost}, the design and motivation of the cost function
    is presented.
    
    \subsection{Simulation set-up}\label{sec:simset}
    The simulations are performed using the numerical tool \DREAM~
    (Disruption Runaway Electron Analysis Model).
    \DREAM~simulates the plasma evolution of tokamak disruptions 
    self-consistently, including the evolution of the plasma temperature, 
    current densities, electron and ion densities, ion charge states, and 
    electric field. 
    The simulations are performed with a multi-fluid plasma model, with 
    the possibility of simulating some, or all, subsets of the total 
    electron population kinetically.
    \DREAM~divides the electrons into up to three populations. 

    Both in the fluid and the kinetic simulations performed here, the 
    bulk and RE populations are treated as distinct fluids.
    The bulk electrons are characterized by their density $\ncold$, 
    temperature $\Tcold$ and the Ohmic current density $\johm$. 
    The REs are described by their density $\nre$, which also yields the 
    runaway current density $\jre=ec\nre$, because they are assumed to 
    move with the speed of light parallel to the magnetic field. 
    In reactor-scale tokamak disruptions, this approximation, i.e. 
    $v_{\rm re}=c$, is typically valid, as supported by \citep{REspeedc}.
    The total current density $\johm + \jre$ is governed by the evolution 
    of the poloidal flux $\psi(r)$ \citep{dream}. 
    For the electrical conductivity $\sigma$ which enters in this 
    evolution, we use the model described by \citet{Redl}, that is 
    valid for arbitrary plasma shaping and collisionality, and takes into 
    account the effects of trapping. 
    Neutral collisions -- which are only relevant at very low post-TQ 
    temperatures that yield strongly unfavourable outcomes -- are not 
    accounted for in the resistivity, as they are not relevant in the 
    parameter ranges of interest \citep{fluidSources}.

    In the kinetic simulations, there is an additional population of 
    superthermal electrons that are represented by the distribution function
    $\fhot$, which is evolved using a pitch-angle averaged Fokker--Planck 
    equation~\citep{dream}.
    This division of the hot and cold populations is especially suitable 
    when there are two distinct Maxwellian electron populations present in 
    the plasma, as is the case when cold materials are injected into a hot 
    plasma during massive material injection.
    
    The temperature of the cold population is determined by a time 
    dependent energy balance equation, accounting for Ohmic heating, line 
    and recombination radiation and bremsstrahlung, as well as a radial 
    heat transport, according to equation (43) of \citep{dream}.  
    The ionization, recombination and radiation rates are taken from the ADAS 
    database for neon and the AMJUEL database\footnote{Documentation for the AMJUEL database: \href{https://www.eirene.de/Documentation/amjuel.pdf}{www.eirene.de/Documentation/amjuel.pdf}.} for deuterium. 
    AMJUEL contains coefficients which account for opacity to Lyman radiation, 
    which is important at high deuterium densities \citep{twoStage}%
    \footnote{The effective ionization loss rate, including the line radiation 
    emitted during the bound-bound transitions leading up to the ionization, was 
    adjusted to enforce that it only takes values larger than the ionization 
    potential for each ionization event. This adjustment was only active at 
    electron densities $\sim 10^{22}\,\rm m^{-3}$ and temperatures 
    $\lesssim 1\,\rm eV$, close to the limits of the validity range for the 
    provided polynomial fit.}.
    The cold electron density is constrained by the condition of
    quasi-neutrality.
    
    In the kinetic model, the Dreicer and hot-tail generation mechanisms 
    are naturally present as velocity space particle fluxes, while in the 
    fluid model, they are both implemented as generation rates.
    The Dreicer generation rate is computed by a neural network
    \citep{DreicerNN} if the input parameters are within the training 
    interval. 
    However, for low electric fields, outside the training interval 
    ($E/E_{\rm D}<0.01$), the values given by the neural network are inaccurate. 
    In such cases, an analytic extrapolation is performed to recover a 
    similar asymptotic behaviour as to the Connor-Hastie formula \citep{Connor1975}, 
    i.e. 
    $\dd \nre / \dd t \propto \exp(-A (E_{\rm D}/E)^b)\rightarrow0$ as 
    $E_{\rm D}/E\rightarrow\infty$.
    Here, $A$ and $b$ are positive parameters fitted to the 
    dependence of the neural network on the electric field for $E/E_{\rm D} > 0.01$, with the density 
    and temperature fixed at the values of the current time step.
    They are fitted under the constraint that $(\dd \nre / \dd t)_{\rm Dreicer}(E/E_{\rm D})$
    should be once differentiable for all values of $E/E_{\rm D}$. 
    
    The hot-tail generation rate is evaluated from the model distribution 
    function of \cite{Smith2008} and using the formula derived by \citet{hottail}  
    -- it is presented in  Appendix~C.4 of \citep{dream}.
    In both fluid and kinetic models, the avalanche generation is 
    determined by the model by~\citet{avalanche}.
    
    In the fluid simulations, the activated generation sources from 
    tritium decay and Compton scattering are evaluated according to 
    \citet{fluidSources}; in the kinetic simulations we use the sources 
    derived in §~\ref{sec:kinSources}.
    During the TQ, the photon flux is set to
    ${\Gamma_{\rm flux}=10^{18}}/(\rm m^2 s)$
    to be consistent with the prediction of the photon flux at ITER for 
    an H-mode discharge at \SI{15}{MA} and  
    \SI{500}{MW} fusion power \citep{solis}.
    However, with the drop in plasma temperature, the fusion power ceases.
    For this reason, the photon flux after the TQ is decreased by a 
    factor of $10^3$, as was done by \citet{Vallhagen2024}.
    
    The disruption simulations are performed in an ITER-like tokamak
    scenario with either a deuterium plasma of density
    \SI{e20}{m^{-3}} for the non-activated scenarios, or a
    deuterium-tritium plasma with the same total ion density and
    50--\SI{50}{\percent} isotope concentrations.
    In both scenarios, the fuel density is spatially uniform.
    Discharges without tritium are simulated for \SI{200}{ms}, while
    activated discharges are simulated for \SI{400}{ms}, in order to 
    capture the interesting runaway current dynamics in both scenarios.
    
    The magnetic field ${B_0=\SI{5.3}{T}}$ on axis, the resistive wall time 
    $\tau_{\rm w}=\SI{0.5}{s}$, the major radius $R_0=\SI{6}{m}$ and the plasma 
    minor radius $a=\SI{2}{m}$.
    \citet{bayesJPP} evaluated the effective radius of the first toroidally closed 
    conducting structure was evaluated by matching the poloidal magnetic 
    energy to that in simulated ITER discharges, yielding 
    $b=\SI{2.833}{m}$~\citep{Vallhagen2024}.
    However, during disruptions, the plasma can be vertically displaced, 
    resulting in a continuous contraction of the confined region, and a 
    corresponding reduction of the poloidal magnetic field energy reservoir 
    that contributes to runaway generation. 
    In this case a more tightly fit wall ($b=\SI{2.15}{m}$) might be more 
    appropriate. 
    Simulations are therefore performed with both of these values of the 
    conducting wall radius, to bracket a plausible range of magnetic to 
    kinetic energy conversion. 
    Furthermore, the geometry is determined by realistic and radially 
    varying shaping parameters congruent with the Miller model 
    equilibrium~\citep{MillerEqui}, used also by \citet{bayesJPP}.
    
    At $t=0$, the plasma current $\Ip=\SI{15}{MA}$ with current density 
    profile $\hat{j}(r)=[1-(r/a)^2]^{0.41}$, 
    the plasma temperature is parabolic with $T(r=0)=\SI{20}{keV}$, and the 
    plasma is fully ionized. 
    The deuterium and neon of the MMI are assumed to consist of cold atom 
    populations, deposited instantaneously and homogeneously, as was done by \citet{bayesJPP}.
    
    To account for the effect of the stochastic magnetic field during the 
    TQ, we employ spatially and temporally constant magnetic 
    perturbations of $\dBB=\SI{0.5}{\percent}$. 
    Such a value of $\dBB$ is consistent with representative transport 
    levels in MHD simulations of ITER disruptions \citep{Hu_2021}. 
    Although we do not consider different values of $\dBB$ here, we note that 
    lower values tend to reduce conducted heat losses as well as the runaway 
    seed, thereby increasing the region in parameter space that yields tolerable 
    disruptions, while the location of the optimum is less affected, 
    consistent with the findings of \citep{bayesJPP}. 
    Following the Rechester--Rosenbluth model~(\citeyear{RechesterRosenbluth}), 
    this magnetic perturbation leads to a transport of hot electrons, REs and 
    electron heat, with a diffusion coefficient $D\propto \abs{v_\parallel} R_0(\dBB)^2$.
    Here, $\abs{v_\parallel}$ is replaced by ${v_{\rm te}=\sqrt{2\Tcold/m_{\rm e}}}$ 
    (local electron thermal speed) for the heat transport and by the speed of light 
    $c$ for the RE transport, where we assume a parallel streaming of the REs 
    along the perturbed field lines.
    As noted by \citet{bayesJPP}, this approach gives an upper bound on 
    the effect of the RE transport as it does not account for the energy 
    and angular dependence of the RE distribution, as well as 
    finite Larmor radius and orbit width effects. 
    In the kinetic simulations, the parallel velocity dependence for the hot 
    electrons is determined by $\fhot$.
    
    The transport of superthermal and runaway electrons is switched off during 
    the current quench, to signify that the magnetic flux surfaces are reformed.
    We still use a small but finite heat transport corresponding to 
    $\dBB=\SI{0.04}{\percent}$ to avoid the development of non-physical 
    narrow hot Ohmic channels \citep{putvinskii,feher}. 
    As long as such channels do not try to form, this low transport level 
    is subdominant to radiative heat losses at post-TQ temperatures.
    In the absence of a clear, widely accepted definition of the end of the TQ, in the 
    simulations, we define the end of the thermal quench by the mean temperature 
    falling below \SI{20}{\electronvolt}.
    We discard simulations where the TQ condition has not been fulfilled within \SI{20}{ms}, which is a factor of $\sim10$ greater than 
    the anticipated TQ time of $1$–$\SI{2}{ms}$ according to \citep{hollmann_lims}.
    In such cases, we can assume that the TQ was not successfully 
    completed and other complications might arise, such as reheating and unsuccessful
    current quench.

    We note, that the criterion used to switch the energetic electron transport off 
    and reduce the heat transport is a source of uncertainty, as the time it takes 
    for flux surfaces to reform might be delayed somewhat after the temperature has 
    reached its post-TQ value. 
    Clearly, longer windows of runaway loss are expected to reduce the final RE 
    current. 
    However, the presence of a relatively small reformed region may be 
    sufficient to support significant RE currents \citep{Tinguely_2023}.     
    
    \subsection{Cost function}\label{sec:cost}
    The cost function of an optimization problem is critical for the 
    optimization to yield meaningful results and, here, the cost function 
    should reliably quantify the risks associated with tokamak disruptions,
    in the form of a single scalar.
    First, the runaway current should be accounted for in the cost 
    function.
    We will consider two plausible choices for representing the runaway 
    current contribution to the cost function -- the maximum runaway 
    current $\Iremax=\max_{t}\Ire(t)$ and the runaway current at the time 
    when it makes up \SI{95}{\percent} of the plasma current 
    $\Irenf=\Ire(t_{\Ire=0.95\Ip})$, which we will call the 95-percent 
    runaway current.
    To account for the advantages and disadvantages of both options 
    (discussed in Appendix~\ref{sec:costapp}), we choose to use the first 
    occurrence of $\Irenf$ unless $\Iremax$ occurs before $\Irenf$ or if 
    $\Ire(t)<0.95\Ip(t)$ throughout the simulation. 
    This will be called the representative runaway current $\Irerep$. 
    
    Additionally, any remaining Ohmic current at the end of the simulation 
    has the potential of being converted into runaway current, as well as 
    indicating if the termination of the discharge was successful.
    Consequently it should also be included in the cost function. 
    Based on~\citep{Lehnen_talk}, the upper safe operational limit of the 
    runaway current is \SI{150}{kA}, which will also be used as the upper
    safe operational limit for the Ohmic current.
    
    Another indication of the success of the disruption mitigation system 
    is the duration of the CQ (CQ time).
    In this work, the CQ time is evaluated through extrapolation, as 
    $\taucq=(t_{\Iohm=\SI{3}{MA}}-t_{\Iohm=\SI{12}{MA}})/0.6$ if the 
    Ohmic current drops below \SI{3}{MA} during the simulation and, otherwise, 
    $\taucq=(t_{\rm final}- t_{\Iohm=\SI{12}{MA}})/(0.8-\Iohmfin/\Ipin)$.
    Here $t_{\Iohm=\SI{3}{MA}}$ ($t_{\Iohm=\SI{12}{MA}}$) is the time when 
    $\Iohm$ is \SI{20}{\percent} (\SI{80}{\percent}) of the initial plasma 
    current.
    As stated by \citet{hollmann_lims}, its safe operational interval is 
    $[50, 150]\,$ms for \SI{15}{MA} discharges.
    
    Finally, the last figure of merit considered is the transported heat 
    loss fraction $\etac$, meaning the fraction of the initial plasma 
    kinetic energy which has been lost from the plasma due to energy 
    transport.
    It is accounted for using the  Rechester--Rosenbluth transport in 
    the collisionless limit, which means that the radial heat diffusion 
    is caused by parallel streaming of electrons along perturbed field 
    lines~\citep{RechesterRosenbluth}, as detailed in §~\ref{sec:simset}.
    For a safe disruption, the transported heat load fraction should be below 
    \SI{10}{\percent}, according to \citet{hollmann_lims}.
    
    Our cost function $\Lcf(\Irerep, \Iohmfin, \taucq, \etac)$ is designed
    to yield a value below one if all components are within their intervals 
    of safe operation.
    To achieve this, $\Irerep$, $\Iohmfin$ and $\etac$ are 
    normalized to their respective safe operational limits.
    The CQ time is shifted and normalized according to 
    $\abs{\taucq-\SI{100}{ms}}/\SI{50}{ms}$, in order to achieve the same 
    behaviour for both the lower and upper limit of safe operation.
    The four figures of merit can be combined using different orders of 
    $p$-norms, and here we have chosen to use the Euclidean norm ($p=2$).
    In Appendix~\ref{sec:costapp} we discuss the impact of different
    choices for $p$.
    
    The final consideration in the design of our 
    cost function is the relative importance of the components.
    With what we have discussed so far, a disruption with 
    $\Irerep=\SI{120}{kA}$ and $\taucq=\SI{100}{ms}$ would yield the same 
    value of $\Lcf$ as $\Irerep=\SI{0}{kA}$ and $\taucq=\SI{60}{ms}$ (if 
    $\Iohmfin=\SI{50}{kA}$ and $\etac=\SI{3}{\percent}$), whereas the 
    latter would be preferable in an actual disruption.
    More specifically, the relevant consideration for the CQ time is that 
    it is between \SI{50}{ms} and \SI{150}{ms} (and preferably with some margin 
    for uncertainties), not necessarily that it is as close to the middle of 
    this safe interval (i.e. 100 ms) as possible. 
    For the RE current however, it is preferable that it is as close to zero 
    as possible, even if RE currents up to \SI{150}{kA} are tolerable. 
    Furthermore, minimizing $\Irerep$ was in this work deemed more important 
    than minimizing $\etac$. 
    To achieve this order of significance when the figures of merit 
    are within their safety intervals, we introduced non-linear (but smooth) 
    rescaling functions $f_i$, which take the form
    \begin{align}
        f_i&=\begin{cases}
            (x_i)^{k_i}, &\text{if } x_i\leq1,\\
            k_i(x_i-1)+1, &\text{if } x_i\geq1,
        \end{cases}
    \end{align}
    for the currents $k_I=1$ and
    $x_I=I/\SI{150}{kA}$, for the conducted heat load $k_\etac=3$ and
    $x_\etac=\etac/\SI{10}{\percent}$, and for the CQ time
    $k_\taucq=6$ and $x_\taucq=\abs{\taucq-\SI{100}{ms}}/\SI{50}{ms}$.
    Note that this implementation of $f_i$ is differentiable.
    
    The final expression for the cost function can thus be written as 
    \begin{equation}
        \Lcf=\sqrt{(C_\Irerep f_\Irerep)^2+(C_\Iohmfin f_\Iohmfin)^2
            +(C_\etac f_\etac)^2+(C_\taucq f_\taucq)^2}.
    \end{equation}
    Each of the components are weighted equally, with weight $C=0.5$, to 
    ensure values below one for scenarios of safe operation.
    
    For some simulations, the cost function cannot be evaluated -- mainly 
    due to lack of convergence or incomplete thermal quench within 
    \SI{20}{ms}.
    In these cases, the cost function was set to a high value ($\Lcf=75 \gg 1$) 
    in order to decrease the probability of the optimizer exploring the 
    surrounding area further.
    
    \subsection{Optimization set-up}\label{sec:bayesopt} 
    Bayesian optimization is an optimization strategy which uses 
    Bayesian inference to estimate a probability distribution function for 
    the entire objective (or cost) function based on a prior set of samples 
    (input-output pairs).
    This is achieved by assuming that all unknowns of the system to be 
    inferred are random variables.
    
    To make viable predictions for the objective function, an 
    assumption of the nature of its probability distribution is needed.
    For this, Bayesian optimization uses stochastic processes, which are 
    infinite collections of random variables~\citep{bayesoptbook}.
    A common stochastic process used for Bayesian optimization 
    is the Gaussian process (GP), in which the random variables are 
    distributed according to multivariate Gaussian distributions.
    In the context of Bayesian optimization, a GP is 
    specified by a mean function $\mu(x)$, which determines the expected 
    objective function value at any point $x$ of the search space, and a 
    covariance function $\phi(x, x')$, which measures the correlation between 
    points $x$ and $x'$~\citep{bayesoptbook}.
    Consequently, this optimization method not only yields an 
    optimum, but also an estimate of the objective function itself in 
    the form of the mean function $\mu(x)$ as well as its uncertainty.
     
    During the optimization, the optimizer advances by taking samples
    of the objective function and successively updating its sample set
    and subsequently its GP estimate of the probability distribution
    function.
    The strategy for choosing which point to sample next is defined by
    the acquisition function, which uses the GP estimate to determine
    which point has the highest potential of improving the current
    optimum.
    
    In this work, the Bayesian optimization was performed using the Python 
    package by \citet{pybayesopt} which uses GPs. 
    The Matérn kernel with smoothness parameter $\nu=3/2$ 
    was used for the GPs and the expected improvement acquisition 
    function was used as the optimization strategy~\citep{bayesoptbook}.
    
    \section{Disruption mitigation optimization}\label{sec:res} 
    The goal of this section is to compare fluid and kinetic 
    simulations of MMI scenarios for a range of injected material densities.
    Since the RE dynamics are different for discharges with and without
    tritium, we will do this comparison both with and without activated
    sources, in §~\ref{sec:act} and §~\ref{sec:nonact}, respectively. 
    While the main focus will be on the differences between the models, 
    we will also discuss the simulated disruption outcomes, 
    including a comparison of our results with those of 
    \citet{bayesJPP}. 
    We will furthermore comment on the impact of choice of wall radius on 
    the results. 
    
    The code used to produce the results in this section, as well as the 
    optimization data, can be found at Zenodo\footnote{Simulation data for this article: \href{https://doi.org/10.5281/zenodo.10998363}{https://doi.org/10.5281/zenodo.10998363}}.
    
    \subsection{Optimization of non-activated discharges}\label{sec:nonact}
    The non-activated scenarios are characterized by only having the Dreicer and 
    hot-tail generation mechanisms as seed generation sources\footnote{Note, that 
    in the non-activated cases we assume that the gamma flux from the wall -- which 
    might be induced by D-D neutron bombardment -- is sufficiently small that the 
    contribution of the corresponding Compton seed to the final runaway current is 
    negligible compared with that of the non-activated sources.}, which tend to 
    be more susceptible to runaway mitigation measures, as we shall see. 
    For the non-activated scenarios, the optimization was performed on the 
    MMI densities within 
    ${\log(\nD/[\SI{1}{m^{-3}}])\in[20, 22]}$ and 
    ${\log(\nNe/[\SI{1}{m^{-3}}])\in[15, 20]}$ for the fluid model.
    For the kinetic model, the intervals were set as 
    ${\log(\nD/[\SI{1}{m^{-3}}])\in[21, 22]}$ and 
    $\log(\nNe/[\SI{1}{m^{-3}}])\in[17.5, 20]$. 
    The total injected number of atoms of $\sim 10^{25}$ corresponding to the 
    upper end the deuterium density ranges, $10^{22}\,\rm m^{-3}$, is an 
    approximate upper bound on material assimilation using shattered pellet 
    injection at ITER~\citep{Vallhagen2024}.
    When using the fluid (kinetic) model, the optimization was performed 
    using 250 (200) disruption simulations.
    
    \begin{figure}
        \centering
        \begin{overpic}[width=9.24cm]{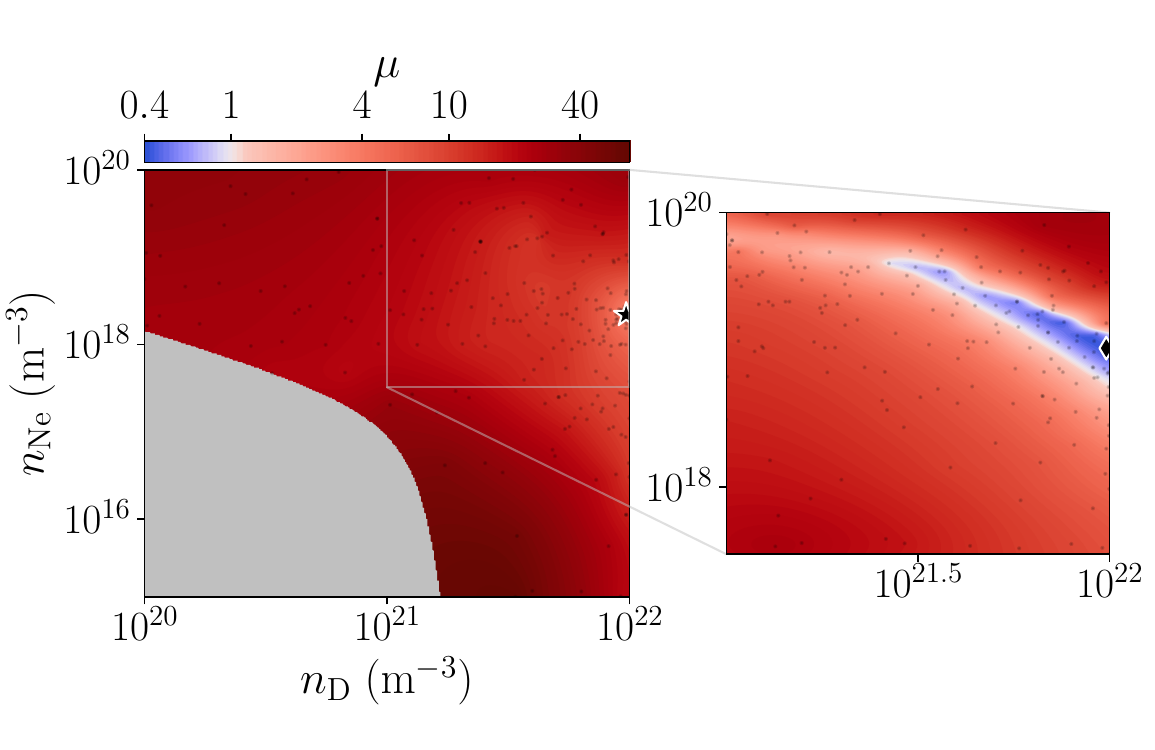} 
            \put(14, 46){\color{white}(a)}
            \put(90, 42){\color{white}(b)}
        \end{overpic}
        \caption{Logarithmic contour plots of the cost function estimate $\mu$
            for the non-activated scenario, generated using (a) the fluid, 
            and (b) the kinetic model in \DREAM.
            Note that the colour mapping is adapted such that blue shades 
            represent regions of safe operation. 
            The black star (diamond) indicates the optimal samples found using 
            the fluid (kinetic) model, while the black dots indicate all optimization 
            samples.
            The grey area covers the region of incomplete TQ.}
        \label{fig:nonactOpt}
    \end{figure}

    Figures \ref{fig:nonactOpt}(a) and \ref{fig:nonactOpt}(b) show the results from the optimization of the
    non-activated scenario for both the fluid and 
    kinetic models, respectively. 
    The most prominent difference between the results of the fluid and kinetic 
    models is that there exists a region of safe operation (as indicated 
    by blue shades in the figure) for the simulations done 
    with the kinetic model, but not with the fluid model. 
    With the kinetic model, the band of safe operation reaches
    from $(\nD, \nNe)=(10^{21.5}, 10^{19.5})\,$\si{m^{-3}} to
    $(\nD, \nNe)=(10^{22}, 10^{19})\,$\si{m^{-3}}.
    Furthermore, the kinetic model yields lower values of the 
    cost function overall compared with the fluid model for 
    ${\log(\nD/[\SI{1}{m^{-3}}])\in[21, 22]}$ and 
    ${\log(\nNe/[\SI{1}{m^{-3}}])\in[17.5, 20]}$.
    
    \begin{figure}
        \centering
        \begin{overpic}[width=10.2cm]{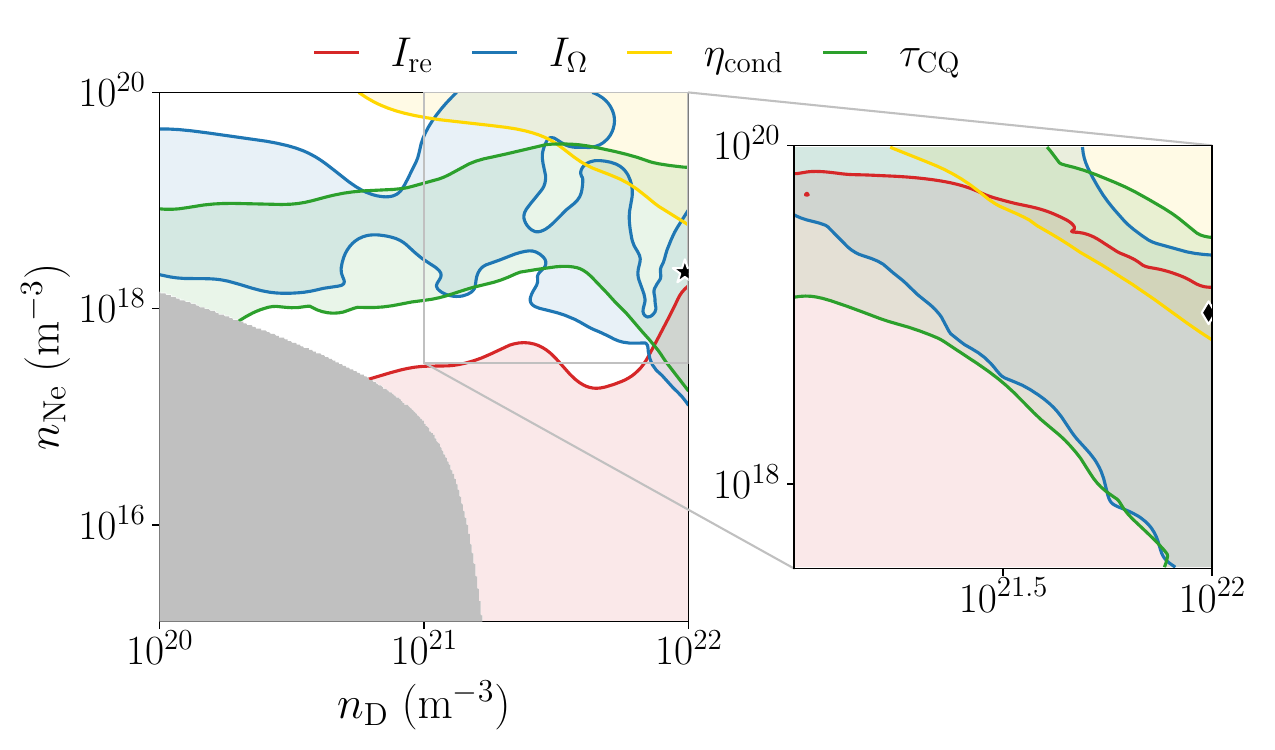}
            \put(13.5, 12.5){(a)}
            \put(63.5, 17){(b)}
        \end{overpic}
        \caption{Regions of safe operation (shaded) for the non-activated
            case with regards to $\Irerep$ (red), $\Iohmfin$ (blue), 
            $\taucq$ (green) and $\etac$ (yellow).
            Panel (a) uses fluid simulations, and panel (b) uses kinetic 
            simulations. 
            The optimal samples are indicated by a star in panel (a) and a
            diamond in panel (b).
            Note that only in the kinetic case, there is a finite 
            intersection of all the regions of safe operation.
            }
        \label{fig:nonactCont}
    \end{figure}
    
    We find that the main difference between the two models is the runaway 
    current. 
    This is illustrated in figure \ref{fig:nonactCont}, where the safe 
    regions of each of the cost function components are indicated by the 
    coloured areas.
    For the fluid model, the safe region of $\Irerep$ only covers the lower 
    part of the ${\log(\nD/[\SI{1}{m^{-3}}])\in[21, 22]}$ and 
    ${\log(\nNe/[\SI{1}{m^{-3}}])\in[17.5, 20]}$ square, but for the kinetic 
    simulations, it covers all but the upper right corner. 
    Thus, with the kinetic model, we find that the band of safe operation 
    that can be observed in figure~\ref{fig:nonactOpt}.b corresponds to the 
    overlap of $\Irerep<\SI{150}{kA}$ and $\etac<\SI{10}{\percent}$, which 
    are the two competing components.
    Similarly, the optimum found using the fluid model is located at a 
    point between the region of safety for $\Irerep$ and $\etac$, since 
    here there is no overlap.
    It is worth noting, however, that such a region of overlap of 
    $\Irerep<\SI{150}{kA}$ and $\etac<\SI{10}{\percent}$
    might not be possible with a tungsten wall, as the additional 
    high-$Z$ impurity can have a more significant negative impact concerning the runaway current
    \citep{Walkowiak2024}, than the improvement the extra radiative losses represent for~$\etac$.  
        
    The shapes of the region of safety for both $\Iohmfin$ and $\taucq$ 
    are also noticeably different, though not to the same degree as
    for $\Irerep$.
    This is simply a consequence of the difference in runaway current.
    In cases when less of the Ohmic current is transformed into runaway current,
    the Ohmic current is higher at the end of the simulation, and 
    the CQ time is therefore longer. 
    This causes both of the regions of $\Iohmfin<\SI{150}{kA}$ and 
    $\SI{50}{ms}<\taucq<\SI{150}{ms}$ to shift upwards when the 
    $\Irerep<\SI{150}{kA}$ region expands.
    The choice of plasma model does not have a significant impact on 
    the conducted heat load -- for both models, the safe region covers
    the upper right corner of the region explored with the kinetic model. 

    \begin{figure}
        \centering
        \begin{overpic}[width=13.44cm]{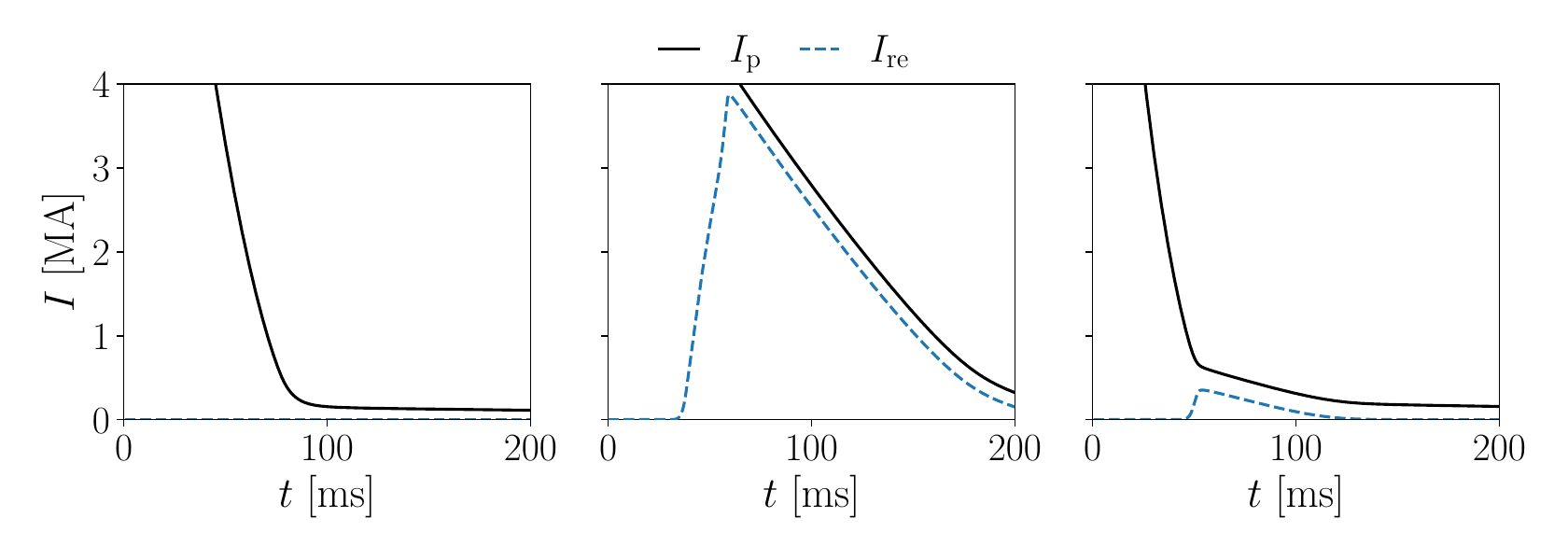}
            \put(30, 27.5){(a)}
            \put(60.5, 27.5){(b)}
            \put(91, 27.5){(c)}
        \end{overpic}
        \caption{Total plasma current (solid) and runaway current 
        (dashed) of the optimum found using the kinetic model, simulated
        with (a) the kinetic model and (b) the fluid model and $a=\SI{2.833}{m}$,
        as well as (c) the fluid model and $a=\SI{2.15}{m}$.}
        \label{fig:nonactCurrs}
    \end{figure}
    
    The results shown in figures \ref{fig:nonactOpt} and \ref{fig:nonactCont} 
    correspond to simulations with a wall radius of \SI{2.833}{m}, but the 
    results are qualitatively similar for a wall radius of \SI{2.15}{m} -- 
    the corresponding landscapes as presented in figure \ref{fig:nonactCont} 
    are much the same.
    The main difference is that with the smaller wall radius, the currents 
    are in general lower and the CQ times are shorter.
    In figure \ref{fig:nonactCurrs}, the plasma and runaway currents are 
    presented for the optimum found using the kinetic model, as well as 
    fluid simulations for the same parameters using $a=\SI{2.833}{m}$ and 
    $a=\SI{2.15}{m}$.
    With the larger wall radius, $\Irerep\approx\SI{4}{MA}$, while the smaller
    wall radius (which also uses a resistive wall time of $0.5\,\rm s$) yields 
    $\Irerep\approx\SI{400}{kA}$. 
    This dramatic reduction in the representative runaway current caused by 
    placing the conducting wall near the plasma, thereby reducing the magnetic 
    energy reservoir for runaway acceleration, is consistent with recent findings 
    reported by \citet{bayesJPP} and \citet{Vallhagen2024}. 
    The reason to consider such a reduced wall radius -- compared with the nominal 
    one matching the total available magnetic energy content in ITER --  is to 
    emulate the reduction of the magnetic energy that is converted to runaway 
    acceleration in a vertical displacement event, as the confined region is 
    gradually scraped off from the plasma channel. 
    This approach can give a lower bound on the representative runaway current. 
    
    The difference in runaway current for the two models is dominated by differences
    in the modelling of the hot-tail RE-generation mechanism, which is highly dependent on
    the energy distribution of the superthermal electrons. 
    In the kinetic simulations, $\fhot$ describes the energy distribution 
    of the superthermal electrons, which enables the hot-tail generation
    to be included in the velocity flux determined by the Fokker-Planck 
    collision operator.
    This energy distribution of the superthermal electrons is not evolved
    in the fluid simulations and, therefore, a model for the energy 
    distribution based on the simulated plasma parameters is needed.
    Such a model, with the purpose of being used to determine the 
    hot-tail generation rate for fluid simulations, was derived by
    \citet{Smith2008}.
    They neglect the electric field and diffusion terms in the Fokker-Planck 
    equation, and instead only retain collisional friction, which gives the 
    solution
    \begin{equation}
        \fhotfl(t,p) = \frac{n_0}{\pi^{3/2}p_{\rm Te}^3}
        \exp\left[-\frac{\left(p^3 + 3\tau(t)\right)^{2/3}}{p_{\rm Te}^{2}}\right],
        \label{eq:f0}
    \end{equation}
    where $n_0$ is the initial free electron density, 
    $p_{\rm Te} = \sqrt{2T_0/m_{\rm e}c^2}$ is the initial thermal momentum of the 
    electrons normalized to $\me c$ (with $T_0$ being the initial electron temperature), 
    and $\tau(t)$ is the time-integrated collision frequency.
    Using this model distribution function for the superthermal electrons,
    the hot-tail generation rate of the fluid model is derived by \citet{hottail} 
    as
    \begin{equation}
        \pdv{n_{\rm re}}{t}=-4\pi {\pcI}^2\pdv{\pcI}{t}\fhotfl(t,\pcI),
        \label{eq:dnrehot}
    \end{equation}
    where the critical momentum $\pcI$ is defined as the momentum at which
    the electric field acceleration and collisional friction balance each
    other; see equation (C.24) in \citep{dream}.\footnote{The correct
    form of equation (C.25) of \citep{dream} is equation \ref{eq:dnrehot}.}

    \begin{figure}
        \centering
        \begin{overpic}[width=9cm]{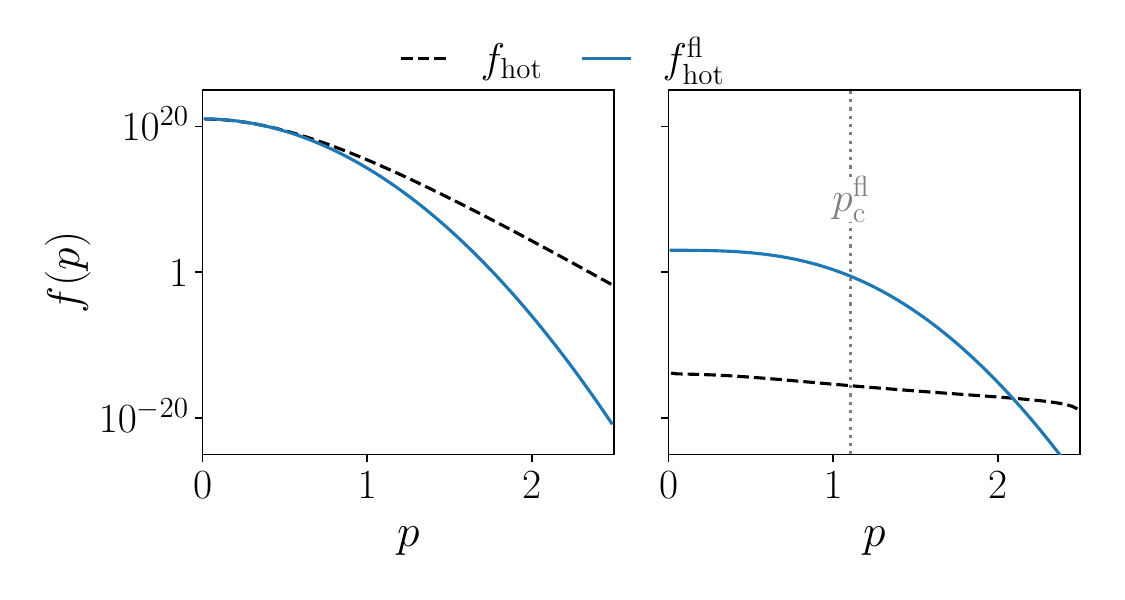}
            \put(39, 41){(a) $t=0$}
            \put(71, 41){(b) ${t=\SI{0.48}{ms}}$}
        \end{overpic}
        \caption{The distribution function for the optimal non-activated case, 
        found using the kinetic model (a) just before and (b) at the 
        end of the TQ.
        The distribution function $\fhot$ (dashed) has been evolved in the kinetic
        simulation and $\fhotfl$ (solid) is the model distribution used in the fluid
        simulation, given by equation \eqref{eq:f0}.
        The latter is based on values of the plasma parameters evolved in the
        fluid simulation.
        }
        \label{fig:nonactDist}
    \end{figure}

    The optimum found using the kinetic model is a clear example of the 
    fluid and kinetic models yielding significantly different RE-current dynamics,
    as shown in figure \ref{fig:nonactCurrs},
    due to the difference in hot-tail generation implementation in the two models.
    For the same MMI densities in a fluid simulation, we obtain a several MA runaway 
    current, where the seed REs are exclusively generated by the hot-tail mechanism, 
    while the kinetic simulation gives practically zero runaway current. 
    In figure \ref{fig:nonactDist}, the distribution function derived by \citet{Smith2008} 
    and the distribution function evolved in the kinetic simulation are plotted 
    just before and at end of the TQ, for $r=\SI{0.85}{m}$ which is the radial 
    position with the highest hot-tail generation rate in the fluid simulation.
    The two distribution functions $\fhot$ and $\fhotfl$ are similar initially,
    as shown in figure \ref{fig:nonactDist}.a. 
    The discrepancy at $p\gtrsim1$ can be fully explained by $\fhot$ being 
    a Maxwell-Jüttner distribution function, while $\fhotfl$ is a non-relativistic 
    Maxwell-Boltzmann distribution. 
    
    For this radius, the hot-tail generation peaks at the end of the TQ in the fluid
    simulation and then the two distribution functions are significantly different, as 
    shown in figure \ref{fig:nonactDist}.b.
    Here $\fhotfl\gg \fhot$ around $\pcI$ (by many orders of magnitude; note the log-scale
    on a wide range), which can be explained by the kinetic simulations accounting 
    for a radial diffusive transport of $\fhot$ due to magnetic perturbations, while 
    the evaluation of $\fhotfl$ does not.
    A generalization of the fluid hot-tail rate, accounting for transport of 
    the hot electrons, would not be straightforward, due to 
    the velocity dependence of the Rechester--Rosenbluth transport.
    The neglect of this transport will lead to a significant overestimation 
    of the runaway seed due to the factor of $\fhotfl(t,\pcI)$ in 
    equation \eqref{eq:dnrehot}.
    This was confirmed by comparing the fluid and kinetic simulations with the radial
    transport removed.

    As shown in figure \ref{fig:nonactOpt}, the optima found with both the 
    fluid and kinetic models are located at high deuterium densities.
    As noted in §~\ref{sec:simset}, at high deuterium densities, the 
    effects of accounting for opacity to Lyman radiation become important.
    When not accounting for opacity to Lyman radiation at these optimum, the
    temperature decays faster, resulting in a significantly larger runaway current, 
    which is in agreement with the conclusions of \citet{twoStage}.
    
    In conclusion, the model distribution function used by the fluid model lacks 
    electric field acceleration and radial transport effects, and it is the transport 
    that dominates the differences observed between the two models. 
    We note, that previous work showed that if the losses due to magnetic 
    perturbations are neglected, i.e.~taking into account all the hot-tail electrons, 
    the final runaway current is overestimated by a factor of approximately four in 
    ASDEX Upgrade \citep{HoppeAUG}, which is in qualitative agreement with our 
    conclusions here.
    
    \subsection{Optimization of activated discharges}\label{sec:act}
    In activated scenarios, the RE seed is expected to be dominated by the generation 
    of REs from tritium beta decay and Compton scattering in cases when the hot-tail and 
    Dreicer generation sources have been successfully mitigated. 
    The same intervals and number of samples were used for the optimization of the 
    activated scenarios as for the non-activated scenarios, 
    except that  the neon density interval was changed to 
    $\log(\nNe/[\SI{1}{m^{-3}}])\in[15, 17.5]$ for the optimization with the kinetic 
    model, in order to include regions of lower runaway current.
    
    \begin{figure}
        \centering
        \begin{overpic}[width=9.24cm]{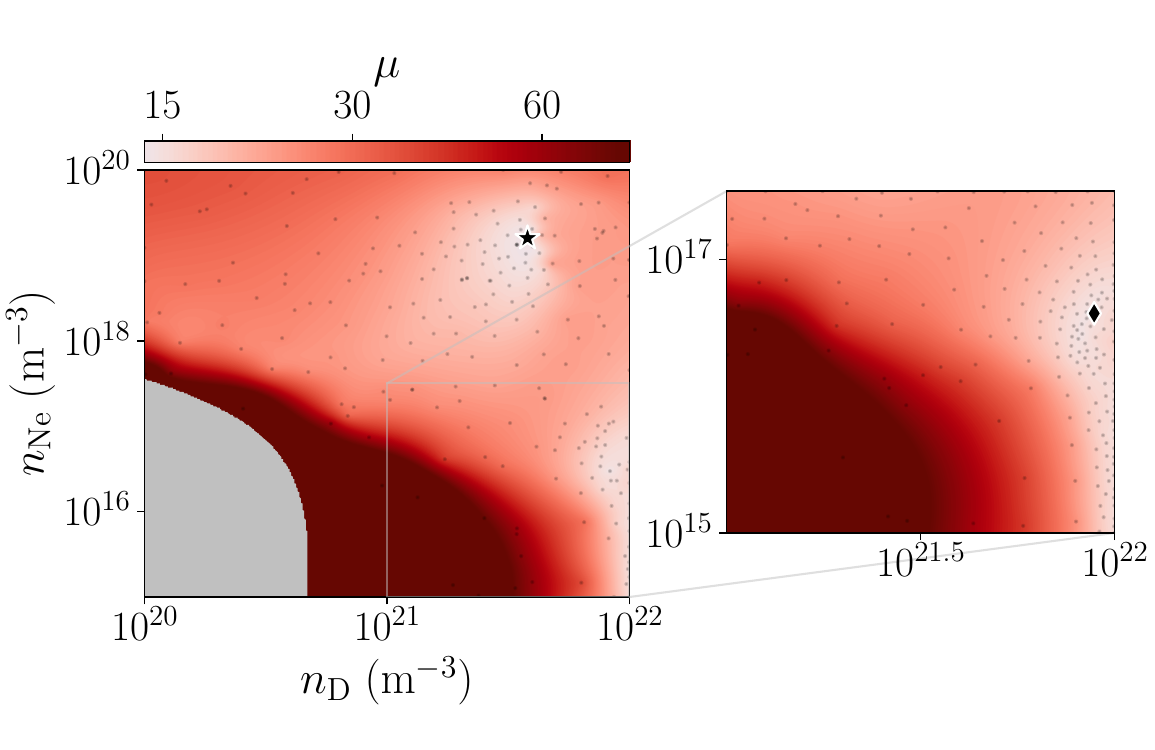} 
            \put(14, 46){\color{black}(a)}
            \put(64, 21){\color{white}(b)}
        \end{overpic}
        \caption{Logarithmic contour plots of the cost function estimate $\mu$
            for the activated scenario, generated using (a) the fluid, 
            and (b) the kinetic model in \DREAM.
            The black star (diamond) indicates the optimal samples found using 
            the fluid (kinetic) model, while the black dots indicate all optimization 
            samples.}
        \label{fig:actOpt}
    \end{figure}

    Figure \ref{fig:actOpt} shows the result from the optimizations of the activated 
    scenarios. 
    There are no regions of safe operation, neither when using the fluid nor
    the kinetic model -- in fact, $\Lcf>10$ (based on the mean prediction $\mu$) 
    everywhere for both models, and both models yield almost identical results.
    From the optimization using the fluid model, there are two local minima, which 
    both are around $\Lcf=15$.
    Since the cost function was designed to distinguish safe from unsafe disruptions,
    and this distinction happens around $\Lcf\sim 1$, it is not as reliable at 
    determining which of the two disruptions is better if their cost function values
    are similar and large, while the individual figures of merit are different. 
    Therefore, the global minimum (in figure \ref{fig:actOpt}.a) is not necessarily 
    safer than the other minimum.
    
    The optimization using the kinetic model presented in figure \ref{fig:actOpt}.b
    has not explored the region around the global minimum found with the fluid model, 
    thus its minimum has the same location as the other local minimum found using the 
    fluid model.
    However, the location of this global optimum did not change when performing a 
    similar optimization using the kinetic model which covered regions 
    surrounding both optima found using the fluid model.
    
    \begin{table}
	\centering
	\caption{Disruption figures of merits for fluid and kinetic simulations for parameters indicated in figure \ref{fig:actCont} with the same markers.}
    \label{tab:act}
	\begin{tabular}{c c c l c c c c c}
		\toprule
		Marker & $\nD$ [m$^{-3}$] & $\nNe$ [m$^{-3}$] & Model & $\Lcf$ & $\Irerep$ [kA] & $\Iohmfin$ [kA] & $\taucq$ [ms] & $\etac$ [\si{\percent}]  \\
		\midrule
        \multirow{2}{1.5cm}{\includegraphics[width=1.5cm]{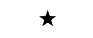}} & \multirow{2}{1.4cm}{$3.8\times 10^{21}$} & \multirow{2}{1.4cm}{$1.6\times 10^{19}$} &  Fluid   & $14.1$  & $4170$ & $103$ & $68.0$  & $23.8$ \\
		& & &Kinetic & $14.7$  & $4320$ & $103$ & $66.3$  & $26.9$  \\
		\midrule
        \multirow{2}{1.5cm}{\hspace{0.1cm}\includegraphics[width=1.3cm]{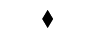}} & \multirow{2}{1.4cm}{$8.9\times 10^{21}$} & \multirow{2}{1.4cm}{$4.0\times 10^{16}$} & Fluid   & $14.3$  & $2790$ & $92.0$ & $201$  & $75.2$   \\
		&  &  & Kinetic & $14.4$  & $2820$ & $88.8$ & $201$  & $75.8$  \\
		\midrule
        \multirow{2}{1.5cm}{\includegraphics[width=1.5cm]{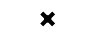}} & \multirow{2}{1.4cm}{$8.2\times 10^{21}$} & \multirow{2}{1.4cm}{$4.9\times 10^{15}$} & Fluid   & $19.4$  & $218$ & $526$ & $408$  & $79.3$   \\
		&  &  & Kinetic & $19.5$  & $225$ & $539$ & $408$  & $79.9$  \\
		\midrule
        \multirow{2}{1.5cm}{\includegraphics[width=1.5cm]{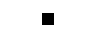}} & \multirow{2}{1.4cm}{$2.7\times 10^{21}$} & \multirow{2}{1.4cm}{$3.5\times 10^{18}$} & Fluid   & $16.5$  & $4030$ & $80.5$ & $108$  & $70.7$   \\
		&  &  & Kinetic & $17.1$  & $4210$ & $83.6$ & $103$  & $71.8$  \\
		\midrule
        \multirow{2}{1.5cm}{\includegraphics[width=1.5cm]{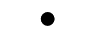}} & \multirow{2}{1.4cm}{$8.8\times 10^{21}$} & \multirow{2}{1.4cm}{$2.1\times 10^{19}$} & Fluid   & $22.7$  & $6810$ & $82.5$ & $43.8$  & $2.55$   \\
		&  &  & Kinetic & $22.8$  & $6830$ & $80.5$ & $43.7$  & $4.73$  \\
		\midrule
        \multirow{2}{1.5cm}{\includegraphics[width=1.5cm]{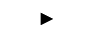}} & \multirow{2}{1.4cm}{$3.0\times 10^{21}$} & \multirow{2}{1.4cm}{$3.4\times 10^{16}$} & Fluid   & $29.3$  & $3960$ & $2200$ & $500$  & $92.9$   \\
		&  &  & Kinetic & $29.4$  & $3950$ & $2200$ & $501$  & $94.0$  \\
	\end{tabular}
\end{table}

    \begin{figure}
        \centering
        \begin{overpic}[width=10.2cm]{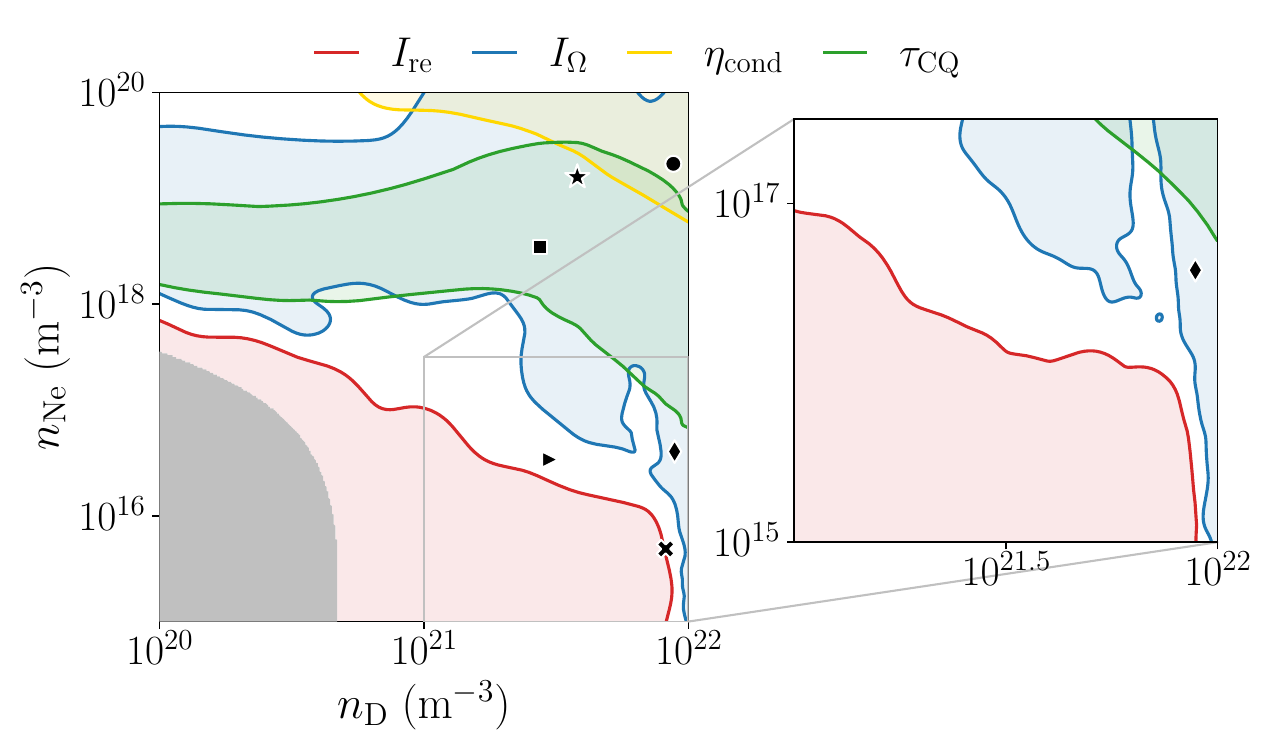}
            \put(13.5, 12.5){(a)}
            \put(63.5, 18.5){(b)}
        \end{overpic}
        \caption{Regions of safe operation (shaded) for the activated
            case with regards to $\Irerep$ (red), $\Iohmfin$ (blue), 
            $\taucq$ (green) and $\etac$ (yellow). 
            Panel (a) uses fluid simulations, and panel (b) uses kinetic 
            simulations. 
            The optimal samples are indicated by a star in panel (a) and 
            a diamond in panel (b).
            The other markers correspond to the cases in table \ref{tab:act}.
            }
        \label{fig:actCont}
    \end{figure}

    That the two plasma models yield very similar results is also clear 
    from figure \ref{fig:actCont}, where the regions of safe operation
    for each figure of merit are plotted.
    The two models yield qualitatively the same results, since the shapes of 
    the safe regions are remarkably alike to the point where it is difficult 
    to distinguish differences due to plasma models from stochastic differences
    of the sampled locations.
    
    In table \ref{tab:act}, fluid and kinetic simulations are compared for 
    different combinations of the deuterium and neon densities, including 
    the two optima (the locations in $\nD$--$\nNe$ space are indicated by 
    symbols in figure \ref{fig:actCont}). 
    The difference between the kinetic and fluid models are in most cases 
    just a few percent, and the relative difference is larger when the 
    figures of merit are small. 
    The largest relative error occurs for the conducted heat load, since
    it is consistently a few percent larger in the kinetic case; this can 
    be explained by the kinetic simulation capturing the transport of 
    the hot electrons.%
    \footnote{Note that for most of the scenarios presented in table \ref{tab:act}, 
    the conducted heat loads are high enough to cause severe wall damage if localized.}

    \begin{figure}
        \centering
        \begin{overpic}[width=9cm]{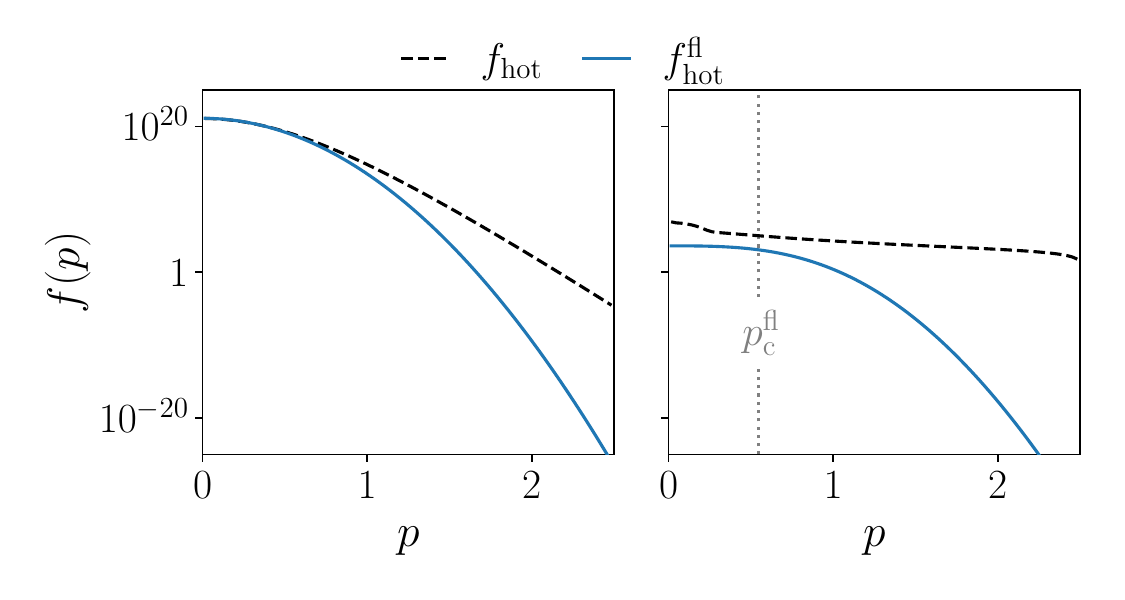} 
            \put(39, 41){(a) $t=0$}
            \put(71, 41){(b) ${t=\SI{0.88}{ms}}$}
        \end{overpic}
        \caption{The distribution function for the optimal activated case found 
        using the fluid model (a) just before and (b) at the end of the TQ.
        The distribution function $\fhot$ (dashed) has been evolved in the kinetic 
        simulation and $\fhotfl$ (solid) is the model distribution used in the fluid
        simulation, given by equation \eqref{eq:f0}.
        The latter is based on values of the plasma parameters evolved in the
        fluid simulation.
        }
        \label{fig:actDist}
    \end{figure}
    
    In absolute terms, only the runaway current varies significantly between the 
    kinetic and fluid results -- for large runaway current cases, the difference 
    can be $\sim\SI{150}{kA}$, while the relative difference is small.
    As shown in table \ref{tab:act}, the kinetic simulations have a higher runaway
    current in all cases but one. 
    The reason for the kinetic model yielding larger runaway current is indirectly
    due to the activated sources.
    The direct generation of electrons above the critical momentum from the activated 
    sources is practically identical for both the fluid and kinetic simulations.
    In the kinetic simulations however, the hot electrons generated by the activated  
    sources with momenta below the critical momentum will cause an increase in RE 
    generation due to flux in velocity space.

    The dynamics of the RE seed generation in the activated scenarios are substantially
    different from those in the non-activated scenarios.
    In the non-activated case, the seed generation was dominated by the hot-tail mechanism. 
    For an activated scenario, $\fhot$ (which includes not only the hot-tail, but also 
    superthermal electrons generated by the Dreicer, Compton scattering and tritium decay 
    mechanisms) will not be depleted due to magnetic perturbations, because there will be a 
    constant generation of hot electrons due to the activated sources.
    This means that the runaway seed generation will not necessarily be overestimated
    with the fluid model, as it was in the non-activated case.
    
    In figure \ref{fig:actDist}, the distribution functions $\fhot$ and $\fhotfl$
    are plotted at the beginning and end of the TQ, at $r=\SI{1.05}{m}$, for the
    optimal activated scenario found using the fluid model.
    We see that, in contrast to the non-activated case, 
    $\fhotfl\ll \fhot$ around $\pcI$ at the end of the TQ for the activated case
    (note the scale of the $y$-axis).
    The fluid simulation should still overestimate the runaway seed generation rate
    due to the neglect of superthermal electron transport in the fluid hot-tail model,
    which should lead to $\fhotfl\gg \fhot$ as in the non-activated cases. 
    However, the fluid model does not capture the additional dynamics in the superthermal 
    momentum range caused by the generation of hot electrons from the 
    activated sources, as discussed above, and this explains why $\fhotfl\ll \fhot$
    in the activated scenarios.
    Furthermore, for the activated cases, the inaccurate runaway seed generation will 
    have a smaller impact on the runaway current evolution since the activated 
    sources, and particularly the Compton source, will constitute a significant 
    fraction of the seed.
    For both optima found in the activated scenarios, the Compton scattering 
    generation mechanism dominates the runaway seed generation by several orders 
    of magnitude.
    
    The physical model and methodology of this study are similar to that of
    \citet{bayesJPP}, but the cost function landscapes and the found optima 
    are significantly different from those presented in that work. 
    The most significant difference compared with the simulations presented by 
    \citet{bayesJPP} is the cost function.
    \citet{bayesJPP} used the following cost function:
    \begin{subequations}
        \begin{align}
            \Lcf_{\rm IP}(\Iremax, \Iohmfin, \taucq,\etac)&=\frac{\Iremax}{\SI{150}{kA}} 
            + \frac{\Iohmfin}{\SI{150}{kA}} + 100\,\theta(\taucq)
            + 10 \frac{\etac}{\SI{10}{\percent}}, \label{eq:costIP_L} \\
            \theta(\taucq)&=  \tilde{\Theta}(\SI{50}{ms} - \taucq) 
            + \tilde{\Theta}(\taucq - \SI{150}{ms}),
        \end{align}%
        \label{eq:costIP}%
    \end{subequations}%
    where $\tilde{\Theta}(\tau)= \frac{1}{2}[1+\tanh (\tau/\SI{3.3}{ms})]$.
    In \eqref{eq:costIP_L}, $\Iremax$ is used instead of $\Irerep$, but this does not have a 
    significant impact on the results. 
    Rather, differences between how $\taucq$ and $\etac$ are treated are responsible for
    most of the difference in the results. 
    
    First, the step function behaviour of $\theta(\taucq)$ limits the 
    optimization of the other parameters to only the region of safe operation
    for $\taucq$.
    All samples with $\taucq\notin[50,\ 150]\, \si{ms}$ will 
    have $\Lcf_{\rm IP}>100$, which is an order of magnitude larger than the 
    optimum found by \citet{bayesJPP}, making it difficult to distinguish 
    the impact of the other components on the cost function value.
    
    Second, the transported heat fraction is weighted one order of magnitude
    higher than the runaway current and final Ohmic current.
    This means that $\etac=\SI{10}{\percent}$ will have the same impact on 
    $\Lcf_{\rm IP}$ as $\Iremax=\SI{1.5}{MA}$, but while $\etac=\SI{10}{\percent}$ 
    is marginally tolerable, $\Iremax=\SI{1.5}{MA}$ is not.
    
    The cost function differences explain the difference in landscape of
    $\Lcf_{\rm IP}$ compared with $\Lcf$ here. 
    Furthermore, it explains why no case with $\Iremax<\SI{4}{MA}$ is found
    by \citet{bayesJPP} -- the cases in table \ref{tab:act} with $\Iremax<\SI{4}{MA}$
    have $\taucq>\SI{150}{ms}$ and $\etac\approx\SI{80}{\percent}$.
    This yields $\Lcf_{\rm IP}\sim200$ which is a rather high value of the cost
    function according to the scale in figure 1 of \citep{bayesJPP}.

    Apart from the difference in cost function landscape, the optimum 
    found using the fluid model is shifted to higher values of $\nNe$ and lower 
    values of $\nD$ compared with the optimum found by \citet{bayesJPP}.
    The main reason for this shift is the decrease of the photon flux from 
    ${\Gamma_{\rm flux}=10^{18}}/(\rm m^2 s)$ to 
    ${\Gamma_{\rm flux}=10^{15}}/(\rm m^2 s)$ after the TQ, which was not accounted for 
    by \citet{bayesJPP}.
    This change in the photon flux leads to a delay in the runaway current ramp up, 
    resulting in a longer decay time for the Ohmic current.
    Therefore, $\taucq$ is increased from \SI{49}{ms} to \SI{67}{ms}, and 
    thus would have a large impact also on $\Lcf_{\rm IP}$.
    
    \begin{figure}
        \centering
        \includegraphics[width=13.44cm]{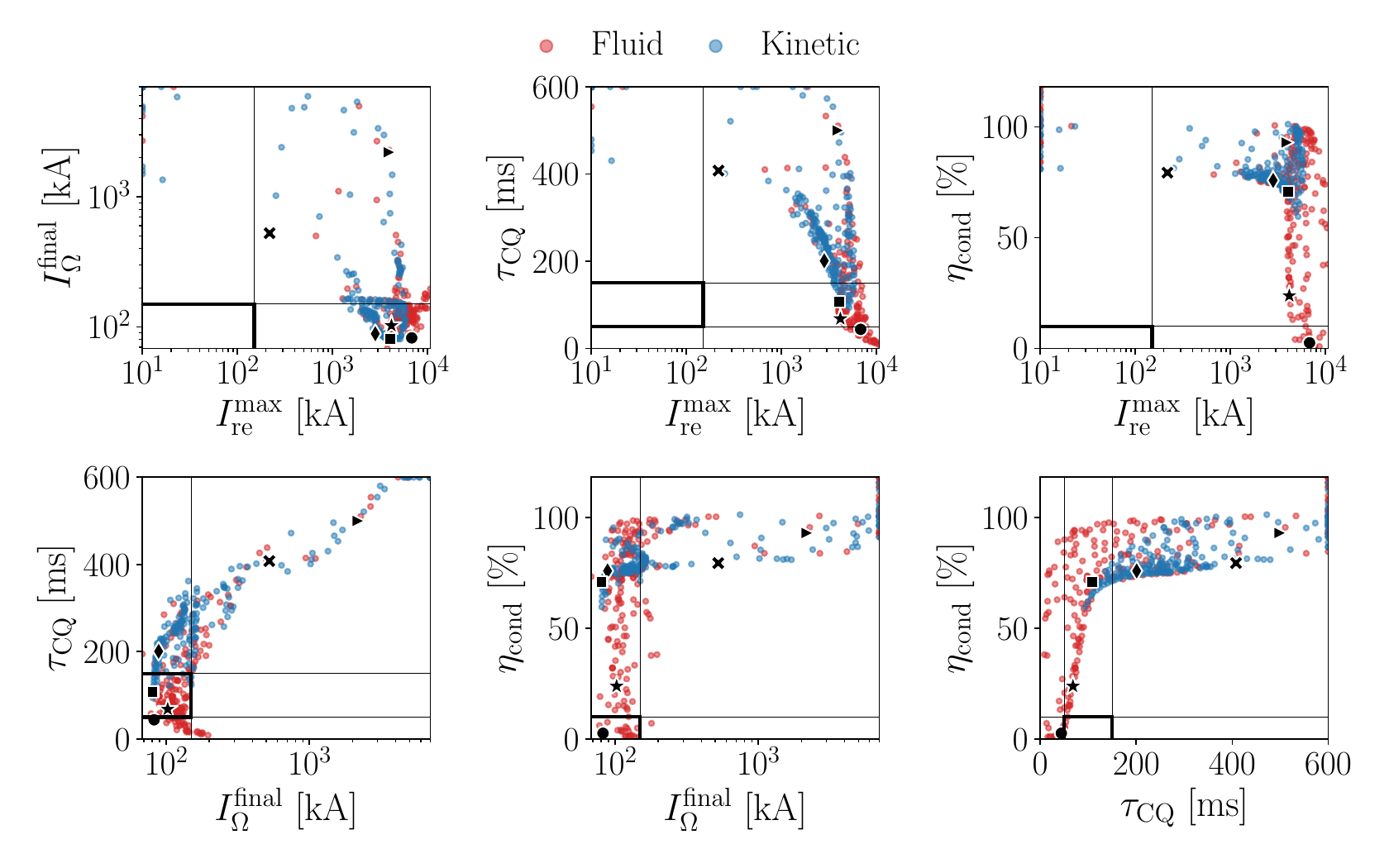}
        \caption{Projections of the simulation dataset to all the two-dimensional
        subspaces of the figure of merit space $(\Irerep, \Iohmfin, \taucq, \etac)$, 
        corresponding to the optimizations of the activated scenario.
        Both kinetic (blue) and fluid (red) simulations are shown.
        The intervals of safe operation for each cost function component
        is indicated by the black lines.
        This figure illustrates the trade-off between the different
        cost function components. 
        The optimal samples found during the optimizations 
        using the fluid (kinetic) model is indicated by a black
        star (diamond), while the other cases in table \ref{tab:act} are
        indicated as black markers of different shapes.
        Samples located outside of the plotted domains are placed at the 
        edges.
        }
        \label{fig:actCloud}
    \end{figure}

    An important similarity between our result and those of \citet{bayesJPP}
    is that we do not find any $\nD$-$\nNe$ combinations yielding tolerable
    or close to tolerable figures of merit in the activated case.
    For the full explored region, $\Lcf\gg1$, and as we mentioned earlier,
    the cost function was not designed to be particularly informative at these values.
    To study correlations and trade-offs between different figures of merit, 
    another approach is needed.
    
    We can use that each simulation which we have performed corresponds to a 
    point in the four dimensional parameter space spanned by our four figures of merit, 
    i.e. $\Irerep, \Iohmfin, \taucq, \etac$.
    In figure \ref{fig:actCloud}, these points have been projected on two-dimensional
    subspaces spanned by pairs of the figures of merit, which illustrates potential 
    trade-offs between each pair.
    Figure \ref{fig:actCloud}(a-c) clarify that for scenarios with $\Irerep<\SI{1}{MA}$, we get
    $\taucq>\SI{400}{ms}$ and $\etac>\SI{75}{\percent}$.
    More specifically, the correlation between $\Irerep$ and $\etac$ suggests
    that it is not possible to achieve moderate values for the runaway current 
    and the transported heat fraction simultaneously -- when $\Irerep<\SI{4}{MA}$, 
    then $\etac>\SI{75}{\percent}$, and {\em vice versa}. 
    This is related to the almost rectangular envelope bounding the point cloud from 
    the side of low values in the $\Irerep$--$\etac$ panel.
    
    Disregarding the runaway current, there are some points of safety overlap for the 
    other three figures of merit; see the lower panels of figure \ref{fig:actCloud}.
    All simulations with $\taucq\in[50,150]\,$ms have $\Iohmfin<\SI{150}{kA}$, and 
    since there is a slight overlap between $\taucq\in[50,150]\,$ms and 
    $\etac<\SI{10}{\percent}$, here $\Iohmfin<\SI{150}{kA}$ as well.
    This can be related back to figure \ref{fig:actCont}, where we see a small region of 
    overlap for safe values of $\taucq$ and $\etac$, as well as the safe region for
    $\taucq$ being almost fully covered by the safe region for $\Iohmfin$.

    Being able to visualize correlations between the cost function components and 
    which regions of the $\nD$-$\nNe$ space correspond to safe regions of
    the figures of merits is among the benefits of using a stochastic and 
    exploratory optimization algorithm such as Bayesian optimization.
    This feature has enabled the exploration of how kinetic and fluid 
    simulations compare for a large variety of scenarios, while simultaneously 
    yielding information on the disruption evolution not just for the most 
    optimal case but for the full region of interest in the $\nD$-$\nNe$ space.
    Such comprehensive comparison or thorough exploration of MMI of deuterium and neon 
    would not have been possible with a more traditional optimization algorithm, 
    such as gradient descent or the comparison of just a few scenarios.

    \section{Conclusions}
    We have compared kinetic and fluid simulations of disruptions with 
    massive material injection for an ITER-like tokamak set-up.
    With this objective in mind, we derived and implemented kinetic runaway electron 
    generation sources for the Compton scattering and tritium beta decay
    mechanisms.
    Furthermore, we investigated how the hot-tail generation compares between the
    two models, both for activated and non-activated scenarios.
    Comprehensive fluid-kinetic comparisons were facilitated using Bayesian optimization
    of the injected densities of deuterium and neon, which enabled exploration of a large
    variety of different disruption scenarios.
    
    We found that while kinetic and fluid disruption simulations representing activated 
    scenarios generated similar results, there were major differences for the non-activated 
    scenarios, and this was caused by differences in the fluid and kinetic hot-tail 
    generation dynamics. 
    Since the kinetic model evolves the distribution function of the superthermal 
    electrons in the simulations, it takes transport of these particles due to 
    magnetic perturbations into account, while the model distribution used to derive 
    the hot-tail generation rate used in the fluid simulations does not. 
    This has a high impact on the non-activated simulations since this transport 
    leads to a depletion of the distribution function, which limits the runaway 
    seed generation in the kinetic simulations.
    
    In the activated case, however, there is a constant generation of 
    superthermal electrons in the kinetic simulations due to the activated sources,
    which acts against the depletion due to transport, and thus makes the difference
    between the fluid and kinetic models less severe.
    Furthermore, the additional generation of runaway electrons from the activated 
    sources reduces the impact of the hot-tail and Dreicer generation on the magnitude 
    of the runaway current during the CQ.
    This limits the significance of the overestimation of the  runaway seed generation 
    due to the neglect of superthermal electron transport in the fluid hot-tail model.
    
    The cost function used for the Bayesian optimization was
    carefully designed, with the feature of safe disruption simulations corresponding
    to cost function values below unity. 
    This had a significant impact on the cost function landscape in the explored
    MMI density region and the optima found during the optimization, when compared
    to a similar optimization done by \citet{bayesJPP}.

    Outstanding issues that remain to be investigated include the effect of vertical 
    displacements events and other instabilities on the runaway electron dynamics.
    Additionally, there might be remnant magnetic perturbations even after the 
    thermal quench, which will affect the final runaway current.
    Finally, the injected material will in reality not be distributed uniformly. 
    The material injection can also be divided into several stages, and recent 
    studies indicate that multi-pellet injections are advantageous for disruption 
    mitigation \citep{Vallhagen2024}.
    The impact of radial inhomogeneities in the assimilated material were, to some
    extent, explored by \citet{bayesJPP}, finding beneficial effects of outward 
    peaking neon and inward peaking deuterium distributions. 
    However, such inhomogeneities might be less consequential for the runaway current
    evolution in the presence of an effective particle transport during the TQ
    \citep{Vallhagen2024}.

    \section*{Acknowledgements} 
    The authors are grateful to Hannes Bergström, Peter Halldestam and 
    Sarah Newton for fruitful discussions.

    \section*{Funding} 
    This work was supported by the Swedish Research Council (Dnr.~
    2022-02862 and 2021-03943), and by the Knut and Alice Wallenberg foundation (Dnr.~2022.0087).
    The work has been carried out within the framework of the EUROfusion 
    Consortium, funded by the European Union via the Euratom Research and 
    Training Programme (Grant Agreement No 101052200 — EUROfusion). 
    Views and opinions expressed are however those of the authors only and 
    do not necessarily reflect those of the European Union or the European 
    Commission. Neither the European Union nor the European Commission can 
    be held responsible for them.
    
    \section*{Declaration of Interests}
    The authors report no conflict of interest.

    \appendix
    
    \section{Cost function details}\label{sec:costapp}
    Here we detail two aspects of the cost function presented in 
    §~\ref{sec:cost} -- the representative runaway current 
    and the impact of choice of $p$-norm.
    
    \subsection{Representative runaway current}
\begin{figure}
        \centering
        \begin{overpic}[width=12cm]{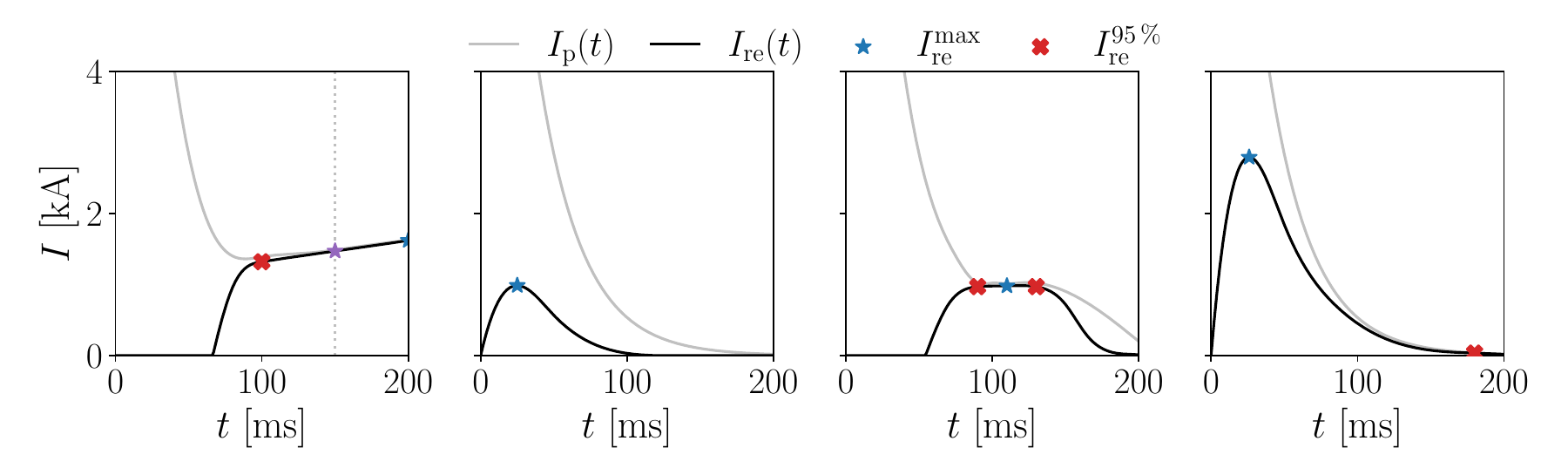}
            \put(22, 23){(a)}
            \put(45.2, 23){(b)}
            \put(68.6, 23){(c)}
            \put(91.5, 23){(d)}
        \end{overpic}
        \caption{(a) A runaway plateau is reached during the disruption, 
            such that the runaway current is slowly increasing and 
            $\Iremax$ depends on the simulation length. If the simulation 
            is stopped at \SI{200}{ms} (\SI{150}{ms}), the blue (purple) 
            star marks $\Iremax$. 
            (b) Criterion for $\Irenf$ is never fulfilled. 
            (c) Criterion for $\Irenf$ is fulfilled more than once. 
            (d) Runaway current peaks significantly early during the 
            simulation, but reaches \SI{95}{\percent} of the total plasma 
            current when both currents are negligible.}
        \label{fig:Ires}
    \end{figure}
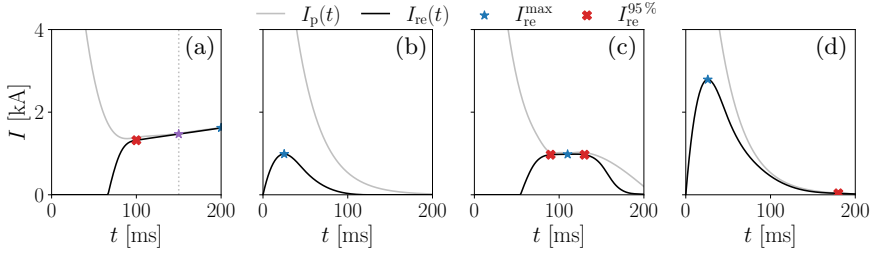

    As described in §~\ref{sec:cost}, we considered two plausible 
    ways to give a representative value for the runaway current; the 
    maximum or the value at the time when $\Ire=0.95\Ip$.
    If a runaway current plateau is reached during the disruption, the 
    runaway current will in most cases be slowly but monotonically 
    increasing with time.
    In this case the maximum runaway current will be dependent on the 
    simulation length, which makes the resulting value arbitrary -- see 
    figure \ref{fig:Ires}.a.
    
    Using the 95-percent runaway current would solve this problem, but it 
    is undetermined when $\Ire(t)<0.95\Ip(t)$ throughout the simulation 
    (figure \ref{fig:Ires}.b) or when $\Irenf$ for several $t$ (figure 
    \ref{fig:Ires}.c). 
    Furthermore, for a hypothetical case where the runaway current near 
    the end of the simulation fulfils $\Ire=0.95\Ip$ while being small, 
    but peaks early during the simulation with an order $\sim\SI{1}{MA}$ 
    without reaching \SI{95}{\percent} of the plasma current, as in figure 
    \ref{fig:Ires}.d, $\Irenf$ would not be a representative value. 
    
    To address all of these shortcomings, we have chosen to use a 
    combination of the maximum and 95-percent current as our 
    representative value for the runaway current. 
    More specifically, we will define 
    $\Irerep=\min_t \{\Iremax, \Irenf\}$.

    \subsection{Choice of p-norm}
    \begin{figure}
        \centering
        \includegraphics[width=10.8cm]{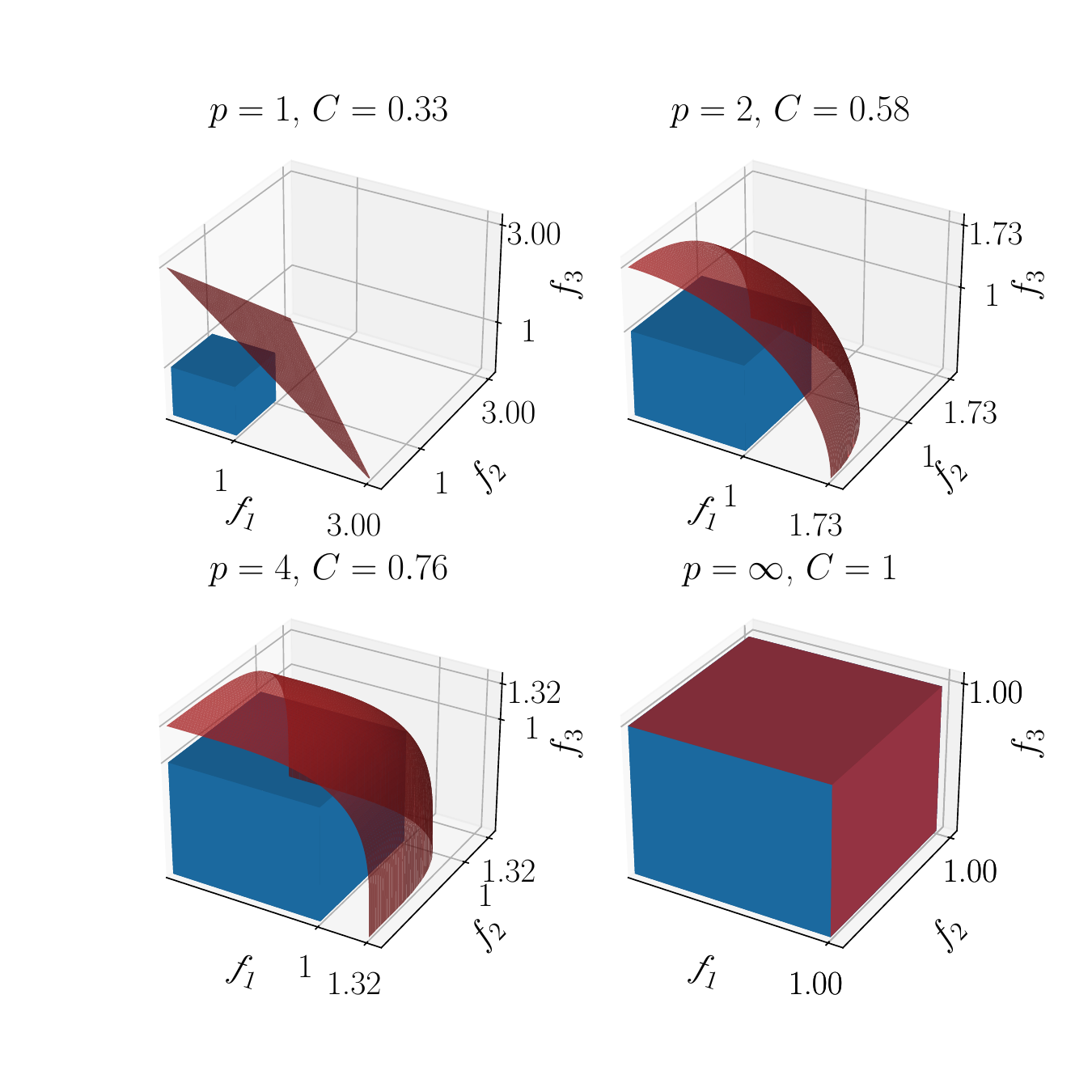}
        \caption{Illustration of the relation between $p$-norm and 
            accuracy of $\Lcf\leq1$ implying safety. 
            For this example, the cost function consists of three 
            components $f_1$, $f_2$ and $f_3$.
            The blue box represents the safe operational region and the 
            red surface implies the $\Lcf=1$ surface. 
            Note that the red surface intersects the axes at $1/C$.}
        \label{fig:unityspheres}
    \end{figure} 
    In §~\ref{sec:cost}, the cost function is described as being 
    designed such that if all components are within their intervals of 
    safe operation the cost function should be smaller than one.
    This was partly achieved by combining the cost function components 
    using the Euclidean norm.
    For safety to be equivalent with the cost function $\Lcf\leq1$, 
    we would need to use the maximum norm instead, i.e. 
    \begin{equation}
        \Lcf=\max{\left(\frac{\Irerep}{\SI{150}{kA}}, 
            \frac{\Iohmfin}{\SI{150}{kA}}, 
            \frac{\etac}{\SI{10}{\percent}}, \frac{\abs{\taucq-\SI{100}{ms}}}{\SI{50}{ms}}\right)}.
        \label{eq:pinf}
    \end{equation}
    However, this would result in the cost function not being once 
    differentiable and only one component would contribute with 
    information about the disruption scenario to the optimizer in any 
    given point.
    
    However, the components could be combined by simply averaging 
    them, i.e.
    \begin{equation}
        \Lcf=\frac14\left(\frac{\Irerep}{\SI{150}{kA}} + 
           \frac{\Iohmfin}{\SI{150}{kA}} + 
           \frac{\etac}{\SI{10}{\percent}} + 
           \frac{\abs{\taucq-\SI{100}{ms}}}{\SI{50}{ms}}\right).
        \label{eq:p1}
    \end{equation}
    While this would imply $\Lcf\leq1$ for all four parameters being 
    inside of their safe operational interval, the opposite implication 
    would not be true, namely $\Lcf\leq1$ implying safety. 
    If there would be a case where only component $x$ is non-zero, this 
    component may rise to four times its safe value with $\Lcf\leq1$ still 
    satisfied.
    
    Both \eqref{eq:pinf} and \eqref{eq:p1} are extreme special cases of 
    the $p$-norm, namely $p=\infty$ and $p=1$, respectively.
    The order of the $p$-norm is thus the trade-off parameter between the 
    largest normalized component dominating the cost function value and 
    thus losing information about the other parameters, and having unity 
    as the exact border between safe and unsafe parameter choices.
    For safety to be implied by $\Lcf\leq1$, the weight of each component must be 
    set as $C=1/(n )^{1/p}$ for $n$ cost function components.
    In figure \ref{fig:unityspheres}, the relation between the value of 
    $p$ and the accuracy of $\Lcf\leq1$ implying safety is illustrated for 
    three arbitrary components $f_1$, $f_2$ and $f_3$.
    We deemed the Euclidean norm ($p=2$) to be close enough to $\Lcf\leq1$ 
    implying safety, without any component being too dominant while large.
    With four components, any component can at most be a factor $1/C=2$ 
    too large for $\Lcf\leq1$, and if $\Lcf\leq0.5$ it is definitely safe.

    \section{Radial current profiles}
\begin{figure}
        \centering
        \begin{overpic}[width=13.44cm]{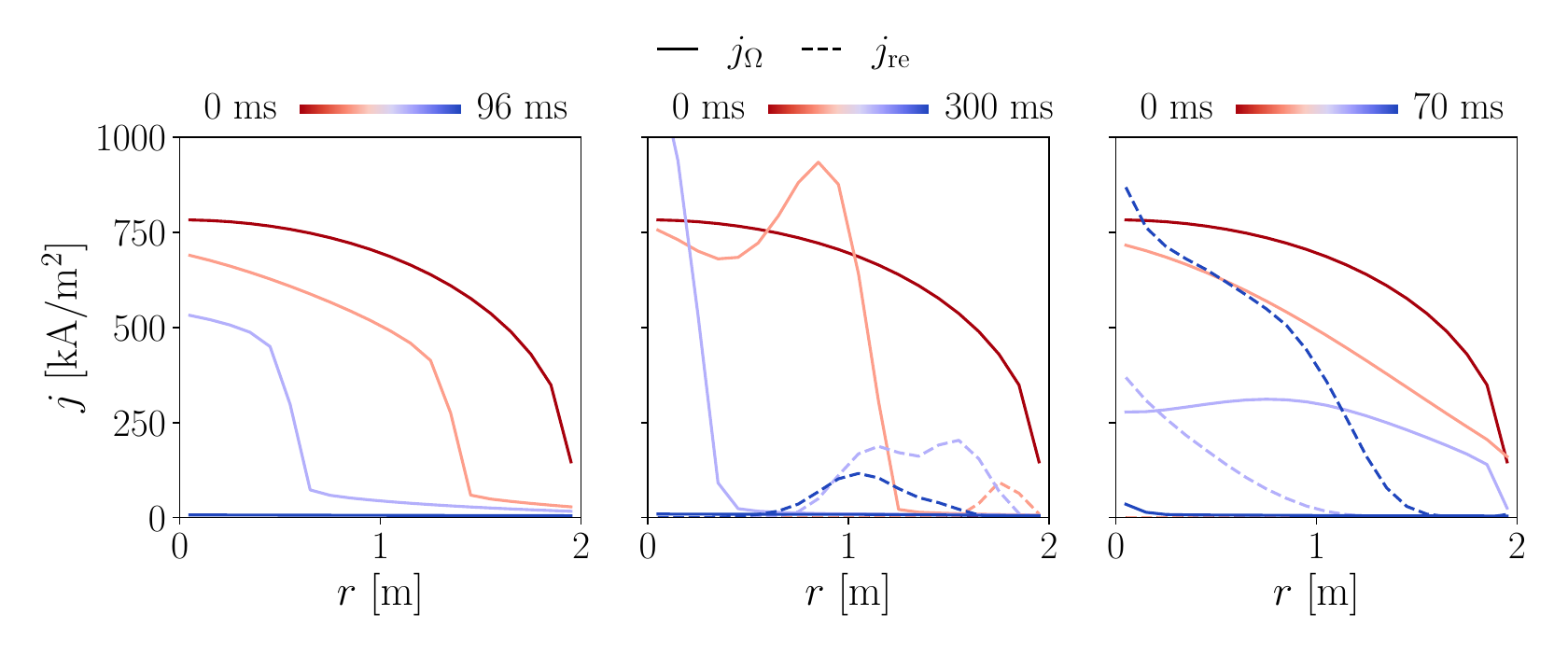}
            \put(33, 30){(a)}
            \put(63, 30){(b)}
            \put(93, 30){(c)}
        \end{overpic}
        \caption{Radial profiles of the current density from kinetic 
        simulations of the optimum (a) for non-activated scenarios, and for 
        activated scenarios found using (b) the fluid model and (c) the 
        kinetic model.
        Both Ohmic (solid) and runaway (dashed) current density profiles
        are shown.
        }
        \label{fig:curprofs}
    \end{figure}

The current density profiles     throughout the simulations of the non-activated and activated optimal     scenarios are presented in figure \ref{fig:curprofs}.
    The Ohmic current density profiles in figure \ref{fig:curprofs}.b are
    non-monotonic or highly peaked in the core, while the runaway current density 
    profiles are hollow.
    In figure \ref{fig:curprofs}.c, especially the runaway current density 
    profiles are strongly peaked in the core.
These features suggest, that some of the scenarios might become prone to MHD instabilities. Thus, to improve the predictive capability of these simulations, evolving the magnetic equilibrium and monitoring MHD stability would be desired. 

    \bibliographystyle{jpp}
    \bibliography{ref}

\end{document}